\documentclass[12pt]{article}
\usepackage{graphicx}
\begin{document}
\newcommand{\beq}{\begin{equation}}
\newcommand{\eeq}{\end{equation}}
\newcommand{\beqa}{\begin{eqnarray}}
\newcommand{\eeqa}{\end{eqnarray}}
\newcommand{\beqar}{\begin{eqnarray*}}
\newcommand{\eeqar}{\end{eqnarray*}}
\newcommand{\al}{\alpha}
\newcommand{\be}{\beta}
\newcommand{\del}{\delta}
\newcommand{\D}{\Delta}
\newcommand{\eps}{\epsilon}
\newcommand{\ga}{\gamma}
\newcommand{\Ga}{\Gamma}
\newcommand{\ka}{\kappa}
\newcommand{\nn}{\nonumber}
\newcommand{\inn}{\!\cdot\!}
\newcommand{\h}{\eta}
\newcommand{\ii}{\iota}
\newcommand{\kk}{\varphi}
\newcommand\F{{}_3F_2}
\newcommand{\la}{\lambda}
\newcommand{\La}{\Lambda}
\newcommand{\na}{\prt}
\newcommand{\Om}{\Omega}
\newcommand{\om}{\omega}
\newcommand{\p}{\phi}
\newcommand{\sig}{\sigma}
\renewcommand{\t}{\theta}
\newcommand{\z}{\zeta}
\newcommand{\ssc}{\scriptscriptstyle}
\newcommand{\eg}{{\it e.g.,}\ }
\newcommand{\ie}{{\it i.e.,}\ }
\newcommand{\labell}[1]{\label{#1}} 
\newcommand{\reef}[1]{(\ref{#1})}
\newcommand\prt{\partial}
\newcommand\veps{\varepsilon}
\newcommand\vp{\varphi}
\newcommand\ls{\ell_s}
\newcommand\cF{{\cal F}}
\newcommand\cA{{\cal A}}
\newcommand\cS{{\cal S}}
\newcommand\cT{{\cal T}}
\newcommand\cC{{\cal C}}
\newcommand\cL{{\cal L}}
\newcommand\cM{{\cal M}}
\newcommand\cN{{\cal N}}
\newcommand\cG{{\cal G}}
\newcommand\cI{{\cal I}}
\newcommand\cJ{{\cal J}}
\newcommand\cl{{\iota}}
\newcommand\cP{{\cal P}}
\newcommand\cQ{{\cal Q}}
\newcommand\cg{{\it g}}
\newcommand\cR{{\cal R}}
\newcommand\cB{{\cal B}}
\newcommand\cO{{\cal O}}
\newcommand\tcO{{\tilde {{\cal O}}}}
\newcommand\bz{\bar{z}}
\newcommand\bw{\bar{w}}
\newcommand\hF{\hat{F}}
\newcommand\hA{\hat{A}}
\newcommand\hT{\hat{T}}
\newcommand\htau{\hat{\tau}}
\newcommand\hD{\hat{D}}
\newcommand\hf{\hat{f}}
\newcommand\hg{\hat{g}}
\newcommand\hp{\hat{\phi}}
\newcommand\hi{\hat{i}}
\newcommand\ha{\hat{a}}
\newcommand\hQ{\hat{Q}}
\newcommand\hP{\hat{\Phi}}
\newcommand\hS{\hat{S}}
\newcommand\hX{\hat{X}}
\newcommand\tL{\tilde{\cal L}}
\newcommand\hL{\hat{\cal L}}
\newcommand\tG{{\widetilde G}}
\newcommand\tg{{\widetilde g}}
\newcommand\tphi{{\widetilde \phi}}
\newcommand\tPhi{{\widetilde \Phi}}
\newcommand\te{{\tilde e}}
\newcommand\tk{{\tilde k}}
\newcommand\tf{{\tilde f}}
\newcommand\tF{{\widetilde F}}
\newcommand\tK{{\widetilde K}}
\newcommand\tE{{\widetilde E}}
\newcommand\tpsi{{\tilde \psi}}
\newcommand\tX{{\widetilde X}}
\newcommand\tD{{\widetilde D}}
\newcommand\tO{{\widetilde O}}
\newcommand\tS{{\tilde S}}
\newcommand\tB{{\widetilde B}}
\newcommand\tA{{\widetilde A}}
\newcommand\tT{{\widetilde T}}
\newcommand\tC{{\widetilde C}}
\newcommand\tV{{\widetilde V}}
\newcommand\thF{{\widetilde {\hat {F}}}}
\newcommand\Tr{{\rm Tr}}
\newcommand\tr{{\rm tr}}
\newcommand\STr{{\rm STr}}
\newcommand\hR{\hat{R}}
\newcommand\M[2]{M^{#1}{}_{#2}}
\parskip 0.3cm

\vspace*{1cm}

\begin{center}
{\bf \Large  Higher derivative corrections to WZ action:}\\ \vspace{0.5cm}{\Large One RR, one NSNS and one NS couplings}

\vspace*{1cm}

{ Ali Jalali\footnote{ali.jalali@stumail.ac.ir} and Mohammad R. Garousi\footnote{garousi@um.ac.ir}  }\\
\vspace*{1cm}
{ Department of Physics, Ferdowsi University of Mashhad,\\ P.O. Box 1436, Mashhad, Iran}
\\
\vspace{1cm}

\end{center}

\begin{abstract}
\baselineskip=18pt

In the first part of this paper, we   calculate the disk-level S-matrix elements of one  RR, one NSNS and one NS vertex operators, and show that  they are consistent with the  amplitudes that have been recently found by applying various   Ward identities. We  show that 
  the massless poles of the amplitude at low energy are fully  consistent with the known D-brane couplings at order  $\alpha'^2$ which involve one  RR or   NSNS   and  two NS fields. Subtracting the massless poles, we then find the contact terms of one RR, one NSNS and one NS fields at order $\alpha'^2$. Some of these terms are reproduced by the Taylor expansion and the pull-back of two closed string couplings, some other couplings are reproduced by  linear graviton in the second fundamental form and by the B-field in the gauge field extension  $F\rightarrow F+B$, in one closed and two open string couplings.  
 
In the second part, we   write  all independent covariant contractions of one RR, one NSNS  and one NS  fields with unknown coefficients. We then constrain the couplings to be consistent with the linear T-duality and with the above contact terms. Interestingly, we have found that up to  total derivative terms and Bianchi identities,  these constraints uniquely fix all the unknown coefficients.     

\end{abstract}

\setcounter{page}{0}
\setcounter{footnote}{0}
\newpage

\section{Introduction and Results } \label{intro}
Higher-derivative couplings in superstring theory may be captured from $\alpha'$-expansion of the  corresponding  S-matrix elements \cite{Gross:1986iv,Gross:1986mw} and from exploring the   dualities of the superstring theory \cite{Kikkawa:1984cp}-\cite{Hull:1994ys}.    The dualities can be implemented either on-shell or off-shell. At the on-shell level, they appear in the S-matrix elements as S-dual and  T-dual Ward identities \cite{Garousi:2011we}. These identities establish connections between  different elements of the scattering amplitude of  $n$ supergravitons. Calculating one  element explicitly in the world sheet conformal field theory, then all other elements   may be generated by the Ward identities \cite{Garousi:2012gh,Velni:2012sv}.  At the off-shell level, on the other hand, the dualities may appear as symmetries of the effective actions which   constrain the couplings \cite{Green:1997tv,Myers:1999ps}.

The effective actions of a single D$_p$-brane in superstring theory at long wavelength limit   are given by the Dirac-Born-Infeld (DBI) and the Wess-Zumino (WZ)   actions. In the string frame they are\footnote{Our index convention is that the Greek letters  $(\mu,\nu,\cdots)$ are  the indices of the space-time coordinates, the Latin letters $(a,d,c,\cdots)$ are the world-volume indices and the letters $(i,j,k,\cdots)$ are the normal bundle indices. }
\beqa
S_p&=&S_p^{DBI}+S_p^{WS}\nonumber\\
&=&-T_p\int d^{p+1}x\,e^{-\phi}\sqrt{-\det\left(P[g+B]_{ab} +F_{ab} \right)}+T_{p}\int e^FP[e^{B}C] \labell{DBI}
\eeqa
where $P[\cdots]$ is the pull-back operator which projects the spacetime tensors to the world volume, \eg $P[g]_{ab}=\frac{\prt X^{\mu}}{\prt\sigma^a} \frac{\prt X^{\nu}}{\prt\sigma^b}g_{\mu\nu}=\tG_{ab}$. The dependence of the closed string fields on the transverse coordinates appears in the action via the Taylor expansion \cite{Garousi:1998fg}.  In the literature, there is a factor of $2\pi\alpha'$ in front of gauge field strength $F_{ab}$. We normalize the gauge field to  absorb this factor. With this normalization, the effective action \reef{DBI} is at the leading order of $\alpha'$. The above actions are invariant  under  T-duality transformations \cite{Myers:1999ps} and are reproduced by the leading order terms  of disk-level S-matrix elements. 

The $\alpha'^2$ corrections to the DBI action should include NSNS and NS fields.   The curvature, the second fundamental form and the dilaton corrections  to the DBI action at order $\alpha'^2$ in the string frame have been found in \cite{Bachas:1999um,Garousi:2009dj,Garousi:2011fc} to be 
\beqa 
S_p^{DBI} &\supset&-\frac{\pi^2\alpha'^2T_{p}}{48}\int d^{p+1}x\,e^{-\Phi}\sqrt{-\tG}\bigg[(R_T)_{abcd}(R_T)^{abcd}-2(\cR_T)_{ab}(\cR_T)^{ab}\nonumber\\
&&\qquad\qquad\qquad\qquad\qquad\qquad-(R_N)_{abij}(R_N)^{abij}+2\bar{\cR}_{ij}\bar{\cR}^{ij}\bigg]\labell{RTN}
\eeqa
where  
 the  curvatures $(R_T)_{abcd}$ and $(R_N)^{abij}$  are related to the projections of  the bulk Riemann curvatures into world volume and transverse spaces, and to the   second fundamental form via the Gauss-Codazzi equations, \ie
\beqa
(R_T)_{abcd}&=&R_{abcd}+\delta_{ij}(\Omega_{\ ac}{}^i\Omega_{\ bd}{}^j-\Omega_{\ ad}{}^i\Omega_{\ bc}{}^j)\nonumber\\
(R_N)_{ab}{}^{ ij}&=&R_{ab}{}^{ij}+g^{cd}(\Omega_{\ ac}{}^i\Omega_{\ bd}{}^j-\Omega_{\ ac}{}^j\Omega_{\ bd}{}^i)\labell{RTRN}
\eeqa
The   curvatures $(\cR_T)_{ab}$ and $\bar{\cR}_{ij}$  are related to the   Riemann curvatures, the   second fundamental form and to the dilaton via the following relations:
\beqa
(\cR_T)_{ab}&=& R^c{}_{acb}+\delta_{ij}(\Omega_c{}^c{}^i\Omega_{\ ab}{}^j-\Omega_{\ ca}{}^i\Omega_b{}^c{}^j)+\prt_a\prt_b\Phi\nonumber\\
\bar{\cR}_{ij}&=& R^c{}_{icj}+ \delta_{ik}\delta_{jl}\Omega^{ab}{}^k\Omega_{ab}{}^l+\prt_i\prt_j\Phi\labell{Rij}
\eeqa
where the world volume indices are raised by the inverse of the pull-back metric\footnote{If one includes   the   trace of the second fundamental form $- \Omega_a{}^a {}^i\Omega_b{}^b {}^j$ into the definition of $\bar{\cR}_{ij}$, then the couplings of one closed string and two open strings can be symmetric under both linear T-duality and S-duality \cite{Jalali:2015xca}. However, there are arguments that the D-brane effective action involving gauge field can not be invariant under S-duality for higher gauge fields \cite{Garousi:2015mdg}. Requiring the effective action to be only invariant under the linear T-duality, as we are going to use in this paper,  then   such extension for $\bar{\cR}_{ij}$ is not required.}. 
In static gauge, the second fundamental form   includes the second derivative of the transverse scalar fields, \ie $\Omega_{ab}{}^i=\prt_a\prt_b\phi^i-\tilde{\Gamma}_{ab}{}^c\prt_c\phi^i+\Gamma_{ab}{}^i$. So  action \reef{RTN} includes the couplings of one graviton or dilaton and two scalar fields. All other couplings between one NSNS and two NS fields at order $\alpha'^2$ have been found in \cite{Jalali:2015xca} by requiring  \reef{RTN} to be invariant under linear T-duality and by requiring the couplings   to be consistent  with the corresponding S-matrix element. The couplings in the string frame are \cite{Jalali:2015xca} 
 \beqa
S_p^{DBI} &\!\!\!\!\!\supset\!\!\!\!\!&-\frac{\pi^2\alpha'^2T_{p}}{12}\int d^{p+1}x\,e^{-\Phi}\sqrt{-\tG}\bigg[ \cR_{bd}\big(\prt_{a}F{^{ab}}\prt_{c}F^{cd}-\prt_{a}F_{c}{}^{d}\prt^{c}F^{ab}\big)+
\frac{1}{2}R_{bdce}\prt^{c}F^{ab}\prt^{e}F_{a}{}^{d}\nonumber\\
&& \qquad\qquad\qquad\qquad+\frac{1}{4}\cR_{d}{}^{d}\big(\prt_{a}F^{ab}\prt_{c}F_{b}{}^{c}+\prt_{b}F_a{}^{c}\prt_{c}F{}^{ab}\big)+\Om_{a}{}^{ai}\prt_{d}H_{c}{}^{d}{}{}_{i}\prt_{b}F^{bc}\nonumber\\
&& \qquad\qquad\qquad\qquad-\Om^{bai}\bigg(\prt_{b}F_{a}{}^{c}\prt_{d}H_{c}{}^{d}{}_{i}
+\prt^{d}F_{a}{}^{c}\prt_{i}H_{bcd}-\frac{1}{2}
\prt^{d}F_{a}{}^{c}\prt_{c}H_{bdi}\bigg)\bigg]\labell{DBI2}
\eeqa
 where $\cR_{ab}$ and $\cR_a{}^a$ are given by
\beqa
\cR_{ab}&=&R^c{}_{acb}+\prt_a\prt_b\Phi\nonumber\\
\cR_{a}{}^{a}&=& R^{ab}{}_{ab}+2\prt^a\prt_a\Phi\labell{del}
\eeqa
which are invariant under linear T-duality. Consistency of the couplings \reef{RTN} with the linear T-duality can also fix $(\prt H)^2$ couplings \cite{Garousi:2009dj}, however, higher order couplings at order $\alpha'^2$, \ie $RH^2$ and $H^4$, are  required for  the consistency of the couplings \reef{RTN}  with nonlinear T-duality in which we are not interested in this paper. Such T-dual couplings have been found in \cite{Robbins:2014ara,Garousi:2014oya} for O-plane.  The gauge invariance  of the couplings \reef{DBI2}   requires $F_{ab}$ to be replaced by $\tB_{ab}=F_{ab}+B_{ab}$.

The curvature corrections to  the WS action have been   found in \cite{Green:1996dd,Cheung:1997az,Minasian:1997mm}
 by requiring that the chiral anomaly on the world volume of intersecting D-branes (I-brane) cancels with the anomalous variation of the WS action. At order $\alpha'^2$, this correction involves curvature squared, \ie $C_{p-3}(R_T\wedge R_T-R_N\wedge R_N)$. Consistency of such couplings with linear T-duality, however,  requires many new couplings involving dilaton, B-field and other RR fields \cite{Becker:2010ij,Garousi:2010rn}, as well as open string fields. On the other hand, consistency of the effective action with the S-matrix element of one RR and one NSNS vertex operators, indicates that there is linear curvature correction to the WS action as well \cite{Garousi:2010ki}. The curvature  transforms to dialton and B-field under linear T-duality, hence,  there should be couplings between one RR and one   NSNS field.  Such couplings  in the string frame   have been found to be   
\cite{Garousi:2010ki} 
\beqa
S^{WS}_p &\!\!\!\!\!  \supset \!\!\!\!\!&-\frac{\pi^2\alpha^2T_{p}}{24}\int d^{p+1}x\,\eps^{a_0\cdots a_p}\left(\frac{1}{3!(p+1)!}\prt_a\cF^{(p+4)}_{ia_0\cdots a_pjk}\prt^aH^{ijk}\right.\labell{LTdual}\\
&&\left.\qquad\qquad +\frac{2}{p!}[\frac{1}{2!}\prt_a\cF^{(p+2)}_{ija_1\cdots a_p}(R_N)_{a_0}{}^{aij}+\frac{1}{p+1}\prt_j\cF^{(p+2)}_{ia_0\cdots a_p}\bar{\cR}^{ij}]\right.\nonumber\\
&&\left.\qquad\qquad+\frac{1}{2!(p-1)!}[\prt^a\cF^{(p)}_{ia_2\cdots a_p}\prt^iH_{aa_0a_1}-\frac{1}{p}\prt^i\cF^{(p)}_{a_1a_2\cdots a_p}(\prt^aH_{iaa_0}-\prt^jH_{ija_0})]\right)\nonumber
\eeqa
where $\cF^{(p)}=dC^{p-1}$. The two closed string couplings  are invariant under linear T-duality and are consistent with the S-matrix element of one RR and one NSNS vertex operators at order $\alpha'^2$ \cite{Garousi:2010ki}. This action includes the couplings of one RR and two transverse scalar fields via the definitions of the curvatures $R_N$ and $\bar{\cR}$. It has been shown in \cite{Jalali:2015xca} that the couplings of one RR and two  NS fields in above action\footnote{The coupling of one RR and two scalars $ (\gamma-1) \prt _j\cF^{(p+2)}_{ia_0\cdots a_p }\Om_a{}^{ai}\Om_b{}^{bj}$ has zero S-matrix and is invariant under  linear T-duality and linear S-duality. Hence, such term could not be fixed in    \cite{Jalali:2015xca}. Since the definition of the curvature $\bar{\cR}_{ij}$  in  \cite{Jalali:2015xca} includes the  trace of the second fundamental form $- \Omega_a{}^a {}^i\Omega_b{}^b {}^j$, then we have set $\gamma=0$ in  \cite{Jalali:2015xca} to have the standard couplings in \reef{LTdual}. Since in the present paper, we are going to impose   consistency of the couplings with S-matrix and linear T-duality, the trace term is not required to be included in the definition of  $\bar{\cR}_{ij}$. So consistency of the couplings \reef{LTdual} with the definition of  $\bar{\cR}_{ij}$ in \reef{Rij} requires   $\gamma=1$.} and in the following action\footnote{To simplify  the couplings in    \cite{Jalali:2015xca}, we have used the identity    $p\,\Om_{a_0}{}^{ai}\prt^{b}\tB_{ba_1}\prt_{i}{\cF}^{(p)}_{aa_2a_3\cdots a_p}+\Om_{a}{}^{ai}\prt^{b}\tB_{ba_0}\prt_{i}{\cF}^{(p)}_{a_1a_2\cdots a_p}=\Om_{a_0}{}^{ai}\prt^{b}\tB_{ba}\prt_{i}{\cF}^{(p)}_{a_1a_2\cdots a_p}$. }:  
%
\beqa
S_p^{WS} &\!\!\!\!\!\supset\!\!\!\!\!&\frac{\pi^2\alpha'^2T_{p}}{12}\int d^{p+1}x\epsilon^{a_0a_1\cdots a_p}\bigg[\frac{1}{2!(p-2)!}\prt^{a}\tB_{a_1a_2}\prt_{b}\tB_{a a_0}\prt^{b}{\cF}^{(p-2)}_{a_3a_4\cdots a_p}\labell{CS2f}\\\nn
&&+\frac{1}{2!(p-1)!}\Om^{bai}\prt_{a}\tB_{a_0a_1}\prt_{b}{\cF}^{(p)}_{ia_2a_3\cdots a_p}
-\frac{1}{(p-1)!} \Om_{a_0}{}^{ai} \prt_{a}\tB_{ba_1}\prt^{b}{\cF}^{(p)}_{ia_2a_3\cdots a_p}\\\nonumber
&& 
-\frac{1}{p!}\Om^{bai} \prt_{a}\tB_{ba_0}\prt_{i}{\cF}^{(p)}_{a_1a_2\cdots a_p} 
+\frac{1}{(p-1)!} \Om_{a_0}{}^{ai}\prt^{b}\tB_{ba}\prt_{i}{\cF}^{(p)}_{a_1a_2a_3\cdots a_p}
\bigg]\nonumber
\eeqa
are consistent with the linear T-duality of one closed and two open strings and with the corresponding S-matrix elements.

In this paper, we calculate  the disk-level S-matrix element of one RR, one NSNS and one NS vertex operators and expand it at low energy. At order $\alpha'^2$, the amplitude has massless poles and contact terms. We will show that the massless poles are reproduced by the corresponding Feynman amplitudes resulting from the  couplings in \reef{DBI}, \reef{RTN}, \reef{DBI2}, \reef{LTdual} and \reef{CS2f}.  Some of the contact terms   are reproduced by the corresponding couplings   in the actions \reef{LTdual} and \reef{CS2f}. The remaining contact terms  should be reproduced by new couplings. We then write all contractions of one RR, one NSNS and one NS fields at order $\alpha'^2$ with unknown coefficients. Imposing   consistency of the couplings  with the above contact terms, one can not uniquely fix the coefficients.  However, we impose the constraints that the couplings are consistent with the above contact terms and are invariant under  the linear T-duality. These fix the coefficients uniquely with the following couplings for $\cF^{(p-2)}$:
\beqa
S_p^{WS} &\!\!\!\!\!\supset\!\!\!\!\!&-\frac{\pi^2\alpha'^2T_{p}}{24}\frac{1}{2(p-2)!}\int d^{p+1}x\, \epsilon^{a_0a_1\cdots a_p}\bigg[H_{aba_2} \
\prt^{a}{}\tB_{a_0a_1} \prt^{b}\cF^{(p-2)}_{a_3a_4\cdots a_p}\nn\\
&&-2H_{aa_2}{}_{i}\prt^{a}{}\tB_{a_0a_1}\prt^{i}{\cF}^{(p-2)}_{a_3a_4\cdots a_p}-H_{ia_1a_2}\prt^{a}\tB_{aa_0}\prt^{i}{\cF}^{(p-2)}_{a_3a_4\cdots a_p}\nn\\
&&- H_{ba_1a_2}\prt_{a_0}\tB^{ab}\prt_{a}{\cF}^{(p-2)}_{a_3a_4\cdots a_p}+(p-2)\tB_{a_0a_1}\prt_{a_2}H^{ic}{}_{a_3}{}\prt_{i}{\cF}^{(p-2)}_{ca_4\cdots a_p}\nn\\
&&+(p-2)H^i{}_{a_2a_3}\prt_{a_1}{}\tB_{aa_0}\prt^{a}{\cF}^{(p-2)}_{ia_4\cdots a_p}+2(p-2)H^{b}{}_{a_2a_3}\prt_{a_1}\tB_{aa_0}\prt^{a}{\cF}^{(p-2)}_{ba_4\cdots a_p}\nn\\
&&+H_{ca_1a_2}\prt^{c}\tB_{aa_0}\prt^{a}{\cF}^{(p-2)}_{a_3a_4\cdots a_p}-H_{ba_1a_2}\prt^a\tB_{aa_0}\prt^b\cF^{(p-2)}_{a_3a_4\cdots a_p}\nn\\
&&
+\frac{(p-2)}{2}H_{aa_2a_3}\prt^b\tB_{a_0a_1}\prt^a\cF^{(p-2)}_{ba_4\cdots a_p}+\frac{(p-2)}{3!}\bigg(3\tB_{a_0a_1}\prt_{i}H^{c}{}_{a_2a_3}\prt^{i}{\cF}^{(p-2)}_{ca_4\cdots a_p}\nn\\
&&+3(p-3)\tB_{a_0a_1}\prt^{c}H^i{}_{a_3a_4}\prt_{a_2}{\cF}^{(p-2)}_{ica_5\cdots a_p}-2\tB^a{}_{a_0}\prt_{b}H_{a_1a_2a_3}\prt^{b}{\cF}^{(p-2)}_{aa_4\cdots a_p}\nn\\
&&+4H_{a_1a_2a_3}\prt^{a}\tB_{aa_0}\prt^{c}{\cF}^{(p-2)}_{ca_4\cdots a_p}-4H_{a_1a_2a_3}\prt^{c}\tB^{a}_{}{a_0}\prt_{c}{\cF}^{(p-2)}_{aa_4\cdots a_p}\bigg)\bigg]\labell{cp-3bfaction}
\eeqa
The following couplings for $\cF^{(p)}$: 
\beqa
S_p^{WS} &\!\!\!\!\!\supset\!\!\!\!\!&-\frac{\pi^2\alpha'^2T_{p}}{24}\frac{1}{p!}\int d^{p+1}x\, \epsilon^{a_0a_1\cdots a_p}\bigg[\prt_{j}{\cF}^{(p)}_{a_1a_2\cdots a_p}H_{ai}{}^{j}\Omega_{a_0}{}^{ai}-\frac{p}{2!}\,\prt_{b}{\cF}^{(p)}_{ia_2\cdots a_p}H^{b}{}_{a_0a_1}\Omega_{a}{}^{ai}\nn\\
&&-p\cF^{(p)}_{ia_2\cdots a_p}\Omega_{a_0}{}^{ai}\prt^{b}H_{aba_1}
-\frac{p}{2!}\cF^{(p)}_{ia_2\cdots a_p}\Omega_{a}{}^{ai}\prt^{b}H_{ba_0a_1}+\frac{p}{2!}\cF^{(p)}_{ia_2\cdots a_p}\Omega^{bai}\prt_{b}H_{aa_0a_1}\nn\\
&&+\frac{p(p\!-\!1)(p\!-\!2)}{3!}\cF^{(p)}_{ijaa_4\cdots a_p}\prt^jH_{a_0a_1a_2}\Omega_{a_3}{}^{ai}+p \tB_{a_0a_1}\cR^{ij}\prt_{j}{\cF}^{(p)}_{ia_2\cdots a_p}-2\tB_{aa_0}\cR^{ai}\prt_{i}{\cF}^{(p)}_{a_1a_2\cdots a_p}\nn\\
&&-2p \tB^{ab}\cR_{ba_1}\prt_{a_0}{\cF}^{(p)}_{aa_2\cdots a_p}+2p \tB_{aa_0} R^{ai}{}_{ba_1}\prt^{b}{\cF}^{(p)}_{ia_2\cdots a_p}-2 \tB^{ab}R_{aa_0bi} \prt^{i}{\cF}^{(p)}_{a_1a_2\cdots a_p}\nn\\
&&+p(p-1) \tB_{a_0a_1} R_{aija_2}\prt^{a}{\cF}^{(p)}_{ija_3\cdots a_p}-p(p-1)\tB^{a}{}_{a_0}R^{i}{}_{a_1a_2}{}^{j}\prt_{a}{\cF}^{(p)}_{ija_3\cdots a_p}\nn\\
&&+p \tB_{ab} R^{bi}{}_{a_0a_1}\prt^{a}{\cF}^{(p)}_{ia_2\cdots a_p}
-2p(p-1)\tB^{a}{}_{a_0}R^{aij}{}_{a_1}\prt_{a_2}{\cF}^{(p)}_{ija_3\cdots a_p}\bigg]\labell{p-1hfaction}
\eeqa
The following couplings for $\cF^{(p+2)}$: 
\beqa
 S_p^{WS} &\!\!\!\!\!\supset\!\!\!\!\!&-\frac{\pi^2\alpha'^2T_{p}}{48}\frac{1}{(p+1)!}\int d^{p+1}x\, \epsilon^{a_0a_1\cdots a_p}\bigg[-4p(p+1){\cF}^{(p+2)}_{ijka_2\cdots a_p}R^{bjk}{}_{a_1}\Omega_{ba_0}{}^{i}\labell{act1}\\
&&-p(p+1){\cF}^{(p+2)}_{ijka_2\cdots a_p} R_{a_0a_1}{}^{jk}\Omega_{a}{}^{ai} 
+\frac{p(p+1)}{2!}\tB_{a_0a_1}\prt^{k}H^{aij}\prt_{a}{\cF}^{(p+2)}_{ijka_2\cdots a_p}\nn\\
&&+(p+1)\tB_{a_0}{}^{b}\prt_{b}H^{aij}\prt_{a}{\cF}^{(p+2)}_{ija_1\cdots a_p}
-(p+1)\tB_{a_0}{}^{b}\prt_{a}H^{aij}\prt_{b}{\cF}^{(p+2)}_{ija_1\cdots a_p}\nn\\
&&
+(p+1)\tB_{a_0}{}^{b}\prt^{a}H_{b}{}^{ij}\prt_{a}{\cF}^{(p+2)}_{ija_1\cdots a_p}
+p(p+1)\tB_{a_0}{}^{b}\prt^{k}H_{a_1}{}^{ij}\prt_{b}{\cF}^{(p+2)}_{ijka_2\cdots a_p}\nn\\
&&
-p(p+1)\tB_{a_0}{}^{b}\prt^{k}H_{b}{}^{ij}\prt_{a_1}{\cF}^{(p+2)}_{ijka_2\cdots a_p}-(p+1) H^{bij} \prt_{a}\tB^{a}{}_{a_0}\prt_{b}{\cF}^{(p+2)}_{ija_1a_2}\nn\\
&&+2\tB^{ab}\prt_{b}H_{a}{}^{ci}\prt_{c}{\cF}^{(p+2)}_{ia_0a_1\cdots a_p}
-2\tB^{ab}\prt_{f}H_{a}{}^{fi}\prt_{b}{\cF}^{(p+2)}_{ia_0a_1\cdots a_p}
-\tB^{ab}\prt^{f}H_{ab}{}^{i}\prt_{f}{\cF}^{(p+2)}_{ia_0a_1\cdots a_p}\nn\\
&&
-\tB^{ab}\prt^{j}H_{ab}{}^{i}\prt_{i}{\cF}^{(p+2)}_{ja_0a_1\cdots a_p}
-\tB^{ab}\prt^{j}H_{ab}{}^{i}\prt_{j}{\cF}^{(p+2)}_{ia_0a_1\cdots a_p}
+(p+1)\tB^{ab}\prt_{a_0}H_{a}{}^{ij}\prt_{b}{\cF}^{(p+2)}_{ija_1 \cdots a_p}\bigg]\nn
\eeqa
And the following couplings for $\cF^{(p+4)}$: 
\beqa
S_p^{WS} &\!\!\!\!\!\supset\!\!\!\!\!&\frac{\pi^2\alpha'^2T_{p}}{48}\frac{1}{(p+1)!}\int d^{p+1}x\, \epsilon^{a_0a_1\cdots a_p}\bigg[{\cF}^{(p+4)}_{ijka_0a_1\cdots a_p}\bigg(
\Omega_{a}{}^{ai}\prt_{b}H^{bjk}-\Omega^{abi}\prt_{a}H_{b}{}^{jk}
\bigg)\labell{p+3bpaction}\\\nn
&&
+(p+1)\cF^{(p+4)}_{ijkla_1\cdots a_p}\bigg(\Omega^{c}{}_{a_0}{}^{i}\prt^{l}H_{c}{}^{jk}
- \Omega_{c}{}^{ci}\prt^{l}H_{a_0}{}^{jk}\bigg)+H^{bjk}\Omega_{a}{}^{ai}\prt_{b}{\cF}^{(p+4)}_{ijka_0a_1\cdots a_p}\bigg]
\eeqa
All above couplings are in the string frame. The couplings of one RR $(p+1)$-form, one H-field and one gauge field in which the RR field strength has two or three transverse indices have been already found in \cite{Velni:2013jha}. Using integration by part, we have checked that the corresponding couplings in \reef{act1} are converted to the couplings found in \cite{Velni:2013jha} after using on-shell relations on the gauge field. The on-shell couplings found in \cite{Velni:2013jha} are consistent with the contact terms of the corresponding S-matrix elements, whereas the off-shell  couplings that we have found are consistent with the S-matrix elements and are also invariant under linear T-duality.

The reason for using the invariance under linear T-duality is that two closed and one open string couplings at order $\alpha'^2$ can not be related to one closed and one open string couplings by nonlinear T-duality as there is no such couplings at order $\alpha'^2$. The above couplings, however,  
  may be related to the standard WS couplings $C_{p-3}(R_T\wedge R_T-R_N\wedge R_N)$ under nonlinear T-duality in which we are not interested in this paper.

An outline of this paper is as follow:  In section 2, we explicitly  calculate the S-matrix element 
of one RR, one NSNS and one NS vertex operators. Up to two unknown integrals,  this amplitude has been calculated in \cite{Velni:2013jha} by using the consistency of the couplings with   Ward identities. Our calculation confirms the result in \cite{Velni:2013jha} and produces the two unknown integrals.   In section 3, we expand  the amplitude at low energy and focus on  the terms at order $\alpha'^2$. In this section, we show that  the massless poles are reproduced by the  corresponding   Feynman amplitudes resulting from the  couplings in \reef{DBI}, \reef{RTN}, \reef{DBI2}, \reef{LTdual} and \reef{CS2f}. After subtracting the massless poles, we obtain the contact terms at order $\alpha'^2$. In this section, we show that some of the contact terms are reproduced by the   pull-back operator and the Taylor expansion of the couplings in  \reef{LTdual} and by two closed and one open string couplings in   \reef{CS2f}. After subtracting the above contact terms, we find the contact terms that should be reproduced by new couplings. In section  4, up to total derivative terms, we write all covariant  contractions with unknown coefficients.  We then constrain the couplings to be consistent with  the   contact terms  found in section 3 and to be invariant under the linear T-duality. We find that up to total derivative terms and Bianch identites, these two constraints fix the couplings uniquely to be those in     \reef{cp-3bfaction}, \reef{p-1hfaction}, \reef{act1}  and \reef{p+3bpaction}.

\section{The S-matrix element in string theory}
 The scattering amplitude of one RR $n$-form, one NS-NS and one NS may be given by the following correlation function:
\beqa
\cA&\sim&<V_{RR}^{(-1/2,-3/2)}(\veps_1^{(n)},p_1)V_{NSNS}^{(0,0)}(\veps_3,p_3)V_{NS}^{(0)}(\veps_2,p_2)>\labell{amp23}
\eeqa
where the vertex operators are \cite{Garousi:1996ad}
\beqa
V_{RR}^{(-1/2,-3/2)}&\!\!\!\!\!=\!\!\!\!\!&(P_-H_{1(n)}M_p)^{AB}\int d^2z_1:e^{-\phi(z_1)/2}S_A(z_1)e^{ip_1\cdot X}:e^{-3\phi(\bz_1)/2}S_B(\bz_1)e^{ip_1\cdot D\cdot  X}:\nonumber\\
V_{NSNS}^{(0,0)}&\!\!\!\!\!=\!\!\!\!\!&(\veps_3\inn D)_{\mu\nu}\int d^2z_2:(\prt X^{\mu}+ip_3\inn\psi\psi^{\mu})e^{ip_3\cdot X}:(\prt X^{\nu}+ip_3\inn D\inn\psi\psi^{\nu})e^{ip_3\cdot D\cdot X}:\nonumber\\
V_{NS}^{(0)}&\!\!\!\!\!=\!\!\!\!\!&\veps_2{}_{\mu}\int dx_3:(\prt X^{\mu}+2ip_2\inn\psi\psi^{\mu})e^{2ip_2\cdot X}:
\labell{closedv}
\eeqa
where  the matrix $D^{\mu}_{\nu}$ is diagonal with $+1$ in the world volume directions and $-1$ in the transverse directions.
The indices $A,\ B,\ldots$ are the Dirac spinor indies and $P_-=\frac{1}{2}(1-\gamma_{11})$ is the chiral projection operator. If 1 in the chiral projection $P_-$ produces couplings for $C^{(n)}$, then the $\gamma_{11}$ produces the couplings for $C^{(10-n)}$. Hence, we consider 1 in the chiral projection and
extend the result to all RR potentials.
 The polarization $\veps_3$ is symmetric for graviton/dilaton and is antisymmetric for B-field, and $\veps_2$ is polarization of gauge field or transvers scalars.
 In the RR vertex operator, $H_{1(n)}$ and $M_p$ are
\beqa
H_{1(n)}&=&\frac{1}{n!}\veps_{1\mu_1\cdots\mu_{n}}\gamma^{\mu_1}\cdots\gamma^{\mu_{n}}\nonumber\\
M_p&=&\frac{\pm 1}{(p+1)!} \eps_{a_0 \cdots a_p} \ga^{a_0} \cdots \ga^{a_p}
\eeqa
where $\eps$ is the volume $(p+1)$-form of the $D_p$-brane and $\veps_1$ is the polarization of the RR form. On-shell conditions are $\veps_i.p_i=p_i.\veps_i=p_i.p_i=0$ for $i=1,2,3$.

Using the standard world-sheet propagators,
 one can   calculate the $X$ and $\phi$ correlators in \reef{amp23}. 
To find the correlator of $\psi$, one should use  the Wick-like rule for the correlation function involving an arbitrary number of $\psi$'s and two $S$'s \cite{Liu:2001qa,Garousi:2010bm}.
Combining the gamma matrices coming from the  $\psi$ correlation in Wick-like rule with the gamma matrices in the RR vertex operator, one finds  the amplitude \reef{amp23} has the  following  trace:
 \beqa
 T(n,p,m)& =&(H_{1(n)}M_p)^{AB}(\gamma^{\alpha_1\cdots \alpha_m}C^{-1})_{AB}A_{[\alpha_1\cdots \alpha_m]}\labell{relation1}\\
& =&\frac{1}{n!(p+1)!}\veps_{1\nu_1\cdots \nu_{n}}\eps_{a_0\cdots a_p}A_{[\alpha_1\cdots \alpha_m]}\Tr(\gamma^{\nu_1}\cdots \gamma^{\nu_{n}}\gamma^{a_0}\cdots\gamma^{a_p}\gamma^{\alpha_1\cdots \alpha_m})\nonumber
 \eeqa
where $A_{[\alpha_1\cdots \alpha_m]}$ is an antisymmetric combination of the momenta and  the polarizations of the NS-NS field and the NS field.
The trace \reef{relation1} can be evaluated for specific values of $n$ and $p$. One can verify that the amplitude is non-zero only for $n=p-3$, $n=p-1$, $n=p+1$, $n=p+3$.
  
The explicit calculation of the S-matrix element of the RR
$(p-3)$-form gives the   result in terms of RR potential \cite{Garousi:2012gh}. Combining the result for RR potential with one transverse index and the result for RR potential with no transverse index, one finds the  following amplitude for D$_4$-brane: 
\beqa
{A}^{(p-3)} &\!\!\!\!\!\sim\!\!\!\!\!& \eps_{a_0\cdots a_4}(\tilde{\cF}^{(2)})^{a_0 a_1}\bigg[ {\veps_2}^{a_3} p_3^{a_2} (p_1\inn N \inn \veps_3^A)^{a_4} \cQ
+ {\veps_2}^{a_3} p_3^{a_2} (p_2\inn \veps_3^A)^{a_4}\cQ_2\nonumber\\
&& \qquad\qquad+{\veps_2}^{a_3} p_3^{a_2} (p_3\inn V\inn \veps_3^A)^{a_4}\cQ_1+\frac{1}{4}
 {\veps_2}^{a_2} p_3\inn V\inn p_3 (\veps_3^A)^{a_3 a_4} \cQ_1\nonumber\\
&& \qquad\qquad-\frac{1}{2} p_3^{a_2} p_3\inn   \veps_2  (\veps_3^A)^{a_3 a_4} \cQ_2\bigg]-\eps_{a_0\cdots a_4}(\tilde{\cF}^{(2)})^{a_0i} {\veps_2}^{a_2} p_3^{a_1} {p_3}_i (\veps_3^A)^{a_3 a_4} {\cal Q}\labell{B0a}
\eeqa
where $\tilde{\cF} $ is the linearized RR field strength in momentum space and $V$ $(N)$ is the flat world-volume (transverse space) metric.  For simplicity, the  amplitude is calculated for $p=4$. It can   be extended to arbitrary $p$ by contracting  the extra word volume indices with the RR field strength.
The closed   and   open string channels appear in the integrals $\cQ,\ \cQ_1$ and $ \cQ_2$ integral. The explicit form  of these integrals have been found in \cite{Garousi:2012gh,Velni:2013jha}, \ie
\beqa
\cQ_1&=&\frac{4\big(\bar{z}_1 \bar{z}_2+z_1\left( z_2-\bar{ z}_1+\bar{z}_2\right)+z_2\left(\bar{ z}_1-2\bar{z}_2\right) }{z_{12} z_{31} z_{2 \bar{1}} z_{3 \bar{1}} z_{1\bar{2}} z_{2 \bar{2}} z_{\bar{1} \bar{2}}}K\nonumber\\
\cQ_2&=&\frac{2z_{2\bar{2}}}{z_{12} z_{32} z_{2 \bar{1}} z_{1 \bar{2}} z_{3\bar{2}} z_{\bar{1} \bar{2}}}K\nonumber\\
\cQ&=&\frac{2z_{1\bar{1}}}{z_{12} z_{13} z_{2 \bar{1}} z_{3 \bar{1}} z_{1\bar{2}} z_{\bar{1} \bar{2}}}K\labell{int}
 \eeqa
 where $z_{ij}=z_i-z_j$ and $z_3=x_3$. There is a measure $\int d^2z_1d^2z_2dx_3$ for all the integrals which we have omitted.  The function $K$ is
\beqa
K&=&z_{ 1\bar{1}}^{p_1.D.p_1}|z_{12}|^{2p_1.p_3}|z_{ 1\bar{2}}|^{2p_1.D.p_3}|z_{13}|^{4p_1.p_2}z_{ 2\bar{2}}^{p_3.D.p_3}|z_{23}|^{4p_3.p_2}
\eeqa
 The integrals in \reef{int}, satisfy the following relation:
\beqa
2p_1\inn N\inn p_3 \cQ+p_3\inn V\inn p_3 \cQ_1+2 p_2\inn p_3 \cQ_2&=&0\,.\labell{id2clos1open} 
\eeqa
The amplitude \reef{B0a} satisfies   the  Ward identity associated with B-field after using the above relation \cite{Garousi:2012gh}.

The   amplitude \reef{B0a}, however,   does not satisfy the Ward identity corresponding to the T-duality. It has been shown in \cite{Velni:2013jha} that the consistency of the amplitude \reef{B0a} with T-dual and gauge symmetry Ward identities requires the following   amplitude for  the RR $(p-1)$-form potential:
\beqa
{A}^{(p-1)}  &\!\!\!\!\!\sim\!\!\!\!\!&\eps_{a_0\cdots a_3}(\tilde{\cF}^{(3)} )^{{a_0}{a_1}}{}_i \Bigg[{p_3}^i \bigg[-\frac{1}{2} p_3^{{a_2}} {\veps_2}^{{a_3}} {\Tr}[\veps_3^S \inn V]\cQ_1 +p_3^{{a_2}}\cQ_2 \bigg((\veps_2 \inn V\inn \veps_3^S)^{{a_3}}-(\phi \inn N\inn \veps_3^A)^{{a_3}}\bigg)\nonumber\\
&&+\frac{1}{2} p_3\inn N\inn \phi  (\veps_3^A)^{{a_2}{a_3}}\cQ_2- {\veps_2}^{{a_3}} (p_1\inn N\inn \veps_3^S)^{{a_2}}\cQ - {\veps_2}^{{a_3}}(p_2\inn V\inn \veps_3^S)^{{a_2}}\cQ_2\bigg]\nn\\
&&-\frac{1}{2} (\veps_2)^{a_3} p_3\inn V\inn p_3 (\veps_3^S)^{a_2 i} \cQ_1- p_3^{a_2} p_3\inn V\inn \veps_2  (\veps_3^S)^{a_3 i}\cQ_2- \phi ^i p_3^{a_2} (p_3\inn V\inn \veps_3^A)^{a_3}\cQ_1\nonumber\\
&&+\frac{1}{4}  \phi ^i p_3\inn V\inn p_3 (\veps_3^A)^{a_2 a_3}\cQ_1+ (\veps_2)^{a_3} p_3^{a_2} (p_1\inn N\inn \veps_3^S)_i\cQ
-\phi ^i p_3^{a_2} (p_1\inn N\inn \veps_3^A)^{a_3}\cQ \nonumber\\
&&+ {\veps_2}^{a_3} p_3^{a_2} (p_2\inn V\inn\veps
_3^S)_i\cQ_2
+ {\veps_2}^{a_3} p_3^{a_2} (p_3\inn V\inn \veps_3^S)_i\cQ_1- \phi ^i p_3^{a_2} (p_2\inn V\inn \veps_3^A)^{a_3}\cQ_2\Bigg]\nn\\
&&+ \eps_{a_0\cdots a_3} (\tilde{\cF}^{(3)} )^{{a_0} {a_1} {a_2}}\bigg[ \frac{1}{3}\veps_2^ {{a_ 3}}\bigg(3p_ 2\inn V\inn \veps_3^S \inn V\inn p_2 \cQ_ 3 +p_1\inn N\inn \veps_3^S \inn N\inn p_1  \cQ + p_1\inn N\inn \veps_3^S \inn V\inn p_3\cQ_1\nonumber\\
&& +3p_ 2\inn V\inn \veps_3^S \inn V\inn p_3 \cQ_ 4 + 2p_1\inn N\inn \veps_3^S \inn V\inn p_2 \cQ_2 -\frac{1}{2} (p_ 1\inn N\inn p_3 \cQ_ {1}+3p_2\inn p_3 \cQ_4){\Tr}[{\veps_3^S}\inn V]\bigg)
\nonumber\\
&& -\frac{1}{2}p_3\inn V\inn p_3\bigg((\veps_2 \inn V\inn \veps_3^S)^{a_3} - (\phi \inn N\inn \veps_3^A)^{a_3} \bigg)\cQ_4 + p_ 3^{a_ 3}\bigg((p_2\inn V\inn \veps_3^S \inn V\inn \veps_2 +p_ 2\inn V\inn \veps_3^A \inn N\inn \phi) \cQ_3
\nonumber\\
&&  +(p_3\inn V\inn \veps_3^S \inn V\inn \veps_2 +p_ 3\inn V\inn \veps_3^A \inn N\inn \phi) \cQ_4 + \frac {1} {3}  (p_ 1\inn N\inn \veps_3^S \inn V\inn \veps_2+p_1\inn N\inn \veps_3^A \inn N\inn \phi ) \cQ_2\bigg)
\nonumber\\
&&
 -\frac {1}{3} p_3\inn V\inn {\veps_2}\bigg((p_ 1\inn N\inn \veps_3^S)^ {{a_ 3}} \cQ_2+3 (p_ 2\inn V\inn \veps_3^S)^ {{a_ 3}}\cQ_3+\frac{3}{2}p_3^{a_3}{\Tr}[{\veps_3^S}\inn V]\cQ_4\bigg)\labell{Fp}\nn\\
&& -\frac {1}{3} p_3\inn N\inn \phi\bigg( (p_ 1\inn N\inn \veps_3^A)^{a_ 3}\cQ_2+3(p_ 2\inn V\inn \veps_3^A)^{{a_ 3}}\cQ_3+3(p_ 3\inn V\inn \veps_3^A)^{{a_ 3}}\cQ_4\bigg)\bigg]\nonumber\\
&&-\eps_{a_0\cdots a_3} (\tilde{\cF}^{(3)})^{a_0}{}_{ ij} p_3^{{a_1}} {p_3}^i\bigg[\phi ^ j (\veps_3^A)^{{a_2} {a_3}}-2 {\veps_2}^{a_3} (\veps_3^S)^{a_2 j}\bigg]\cQ\labell{ext1}
\eeqa
The amplitude is   for $p=3$.  The consistency with the Ward identities, however, could not fix the form of the integrals $\cQ_3$ and $\cQ_4$. It has been pointed out in \cite{Velni:2013jha} that the explicit form of these integrals should be calculated from S-matrix calculations. 

We have explicitly calculated the amplitude \reef{amp23} for RR $(p-1)$-form and found exactly the   result in \reef{ext1} with the following expressions for the two integrals:
 \beqa
\nonumber
\cQ_3&\!\!\!\!\!=\!\!\!\!\!&\frac{2 \left(x_3 \left(\bar{z}_1+z_1-2 z_2\right)+z_2 \bar{z}_1+z_1 \left(z_2-2 \bar{z}_1\right)\right) \left(x_3 \left(\bar{z}_1-2 \bar{z}_2+z_1\right)-2 z_1 \bar{z}_1+\left(\bar{z}_1+z_1\right) \bar{z}_2\right)}{3 \left(z_1-z_2\right) \left(z_1-\bar{z}_1\right) \left(z_2-\bar{z}_1\right) \left(z_1-\bar{z}_2\right) \left(\bar{z}_1-\bar{z}_2\right) \left(x_3-z_1\right) \left(x_3-z_2\right) \left(x_3-\bar{z}_1\right) \left(x_3-\bar{z}_2\right)}K\nonumber\\
\cQ_4&\!\!\!\!\!=\!\!\!\!\!&\frac{4 \left(\bar{z}_1 \bar{z}_2+z_1 \left(-2 \bar{z}_1+\bar{z}_2+z_2\right)+z_2 \left(\bar{z}_1-2 \bar{z}_2\right)\right)}{3 \left(z_2-z_1\right) \left(z_1-\bar{z}_1\right) \left(z_2-\bar{z}_1\right) \left(z_1-\bar{z}_2\right) \left(\bar{z}_1-\bar{z}_2\right) \left(x_3-z_2\right) \left(x_3-\bar{z}_2\right)}K\labell{int2}
\eeqa
The following relation between  the  integrals    $\cQ_2$, $\cQ_3$, $\cQ_4$   has been found in \cite{Velni:2013jha}:
\beqa
3p_3\inn V\inn p_3 \cQ_4+6p_2\inn p_3\cQ_3+2p_1\inn N\inn p_3\cQ_2&=&0\labell{rel2}
\eeqa
By using $\alpha'$ expansions for the integrals \reef{int2} we have checked it to the first order of $\alpha'$.

Using the relations \reef{id2clos1open} and \reef{rel2}, it has been shown in \cite{Velni:2013jha} that the amplitude \reef{Fp} satisfies the Ward identities corresponding to the gauge symmetries. However, it does not satisfy the Ward identity corresponding to the T-duality. It has been shown in \cite{Velni:2013jha} that the consistency of the amplitude \reef{Fp} with T-dual and gauge symmetry Ward identities requires the following amplitude for the RR $(p+1)$-form potential:
\beqa
{A}^{(p+1)} &\!\!\!\!\sim\!\!\!\!&\eps_{a_0a_1 a_2}\bigg\{(\tilde{\cF}^{(4)})^{{a_0}{a_1}}{}_{ij}\Bigg[\frac{1}{4} p_3\inn V\inn p_3 \bigg(\veps_2^{{a_2}} (\veps_3^A)^{ij}+2 \phi ^i (\veps_3^S)^{{a_2}j}\bigg) \cQ_1-\frac{1}{2}p_3^{{a_2}} p_3\inn V\inn \veps_2  (\veps_3^A)^{ij}\cQ_2\nn\\
&&
\!\!\!\!+  \phi ^j p_3^{{a_2}} (p_1\inn N\inn \veps_3^S)^i\cQ+ \phi ^j p_3^{{a_2}} (p_2\inn V\inn \veps_3^S)^i \cQ_2+ \phi ^j p_3^{{a_2}} (p_3\inn V\inn \veps_3^S)^i\cQ_1+{p_3}_i \bigg[\nonumber\\
&&\!\!\!\! -\frac{1}{2}(p_3)^{{a_2}} \phi _j {\Tr}[\veps_3^S\inn V] \cQ_1+  p_3\inn N\inn \phi  (\veps_3^S)^{{a_2}}{}_{j}\cQ_2+ p_3^{{a_2}}\bigg((\veps_2\inn V\inn \veps_3^A)_j-(\phi \inn N\inn \veps_3^S)_j\bigg)\cQ_2\nonumber\\
 &&\!\!\!\! -\bigg(\phi _j (p_1\inn N\inn \veps_3^S)^{{a_2}}-{\veps_2}^{{a_2}} (p_1\inn N\inn \veps_3^A)_j\bigg)\cQ-\bigg(\phi _j (p_2\inn V\inn \veps_3^S)^{a_2}-{\veps_2}^{{a_2}} (p_2\inn V\inn \veps_3^A)_j\bigg)\cQ_2\bigg]\Bigg]\nn\\
&&\!\!\!\! -(\tilde{\cF}^{(4)})^{{a_0}}{}_{ijk}  p_3^{{a_1}} {p_3}^i  \Bigg[{\veps_2}^{{a_2}} (\veps_3^A)^{jk}+2 \phi ^j (\veps_3^S)^{{a_2}k}\bigg]\cQ +(\tilde{\cF}^{(4)} )^{{a_0} {a_1} {a_2}}{}_{i}\bigg[p_{3}{}^i\Bigg(\frac{1}{2}p_3\inn N \inn\phi {\Tr}[{\veps_3^S}\inn V]\cQ_4 \nn\\
&&\!\!\!\!+\frac{1}{3}\bigg(p_1 \inn N\inn \veps_3^A\inn V\inn\veps_2 +p_1\inn N\inn\veps_3^S\inn N\inn \phi   \bigg)\cQ_2+
\bigg(p_2 \inn V\inn \veps_3^A\inn V\inn\veps_2 +p_2\inn V\inn\veps_3^S\inn N\inn \phi   \bigg)\cQ_3\Bigg)
\nn\\
&&\!\!\!\!+\frac{1}{3}\phi^ {i}\Bigg(3p_ 2\inn V\inn \veps_3^S \inn V\inn p_2 \cQ_ 3 +p_1\inn N\inn \veps_3^S \inn N\inn p_1  \cQ + p_1\inn N\inn \veps_3^S \inn V\inn p_3 \cQ_1+ 2p_1\inn N\inn \veps_3^S \inn V\inn p_2 \cQ_2\nonumber\\
&&\!\!\!\!  + 3p_2\inn V\inn \veps_3^S \inn V\inn p_3 \cQ_4 -\frac{1}{2} (p_ 1\inn N\inn p_3 \cQ_ {1}+3p_2\inn p_3 \cQ_4){\Tr}[{\veps_3^S}\inn V]\bigg) -\frac {1}{3} p_3\inn V\inn {\veps_2}\bigg((p_ 1\inn N\inn \veps_3^A)^ {i} \cQ_2\nonumber\\
&&\!\!\!\!+3 (p_ 2\inn V\inn \veps_3^A)^ {i}\cQ_3\bigg)-\frac{1}{2}p_3\inn V\inn p_3\bigg((\veps_2 \inn V\inn \veps_3^A)^{i} - (\phi \inn N\inn \veps_3^S)^{i} \bigg)\cQ_4-\frac{1}{3} p_3\inn N\inn \phi \bigg((p_ 1\inn N\inn \veps_3^S)^{i} \cQ_2\nonumber\\
&& \!\!\!\!+3(p_ 3\inn V\inn \veps_3^S)^{i} \cQ_4+3(p_ 2\inn V\inn \veps_3^S)^{i})\cQ_3\bigg)\cQ\Bigg)\Bigg]\bigg\}\labell{FF}
\eeqa
The amplitude is  for $p=2$. The T-duality could not fix the integrals $\cQ_3$, $\cQ_4$.

We have explicitly calculated the amplitude \reef{amp23} for RR $(p+1)$-form and found exactly the   result \reef{FF} with the explicit form \reef{int2} for   the integrals $\cQ_3$, $\cQ_4$ that we have found in this paper.

Finally, the consistency of the amplitude \reef{FF} with T-dual and gauge symmetry Ward identities requires the following amplitude for the RR $(p+3)$-form potential \cite{Velni:2013jha}:
 \beqa
{A}^{(p+3)} &\sim& \eps_{a_0  a_1}\bigg\{(\tilde{\cF}^{(5)})^{{a_0}{a_1}ijk} \frac{1}{4}\Bigg[\phi _k p_3\inn V\inn p_3 (\veps_3^A)_{ij}\cQ_1+2p_3{}_i\bigg(p_3 \inn N\inn \phi  (\veps_3^A)_{jk}\cQ_2\nonumber\\
&&-2 \phi _j (p_1\inn N\inn \veps_3^A)_k\cQ -2 \phi _j (p_2\inn V\inn \veps_3^A)_k\cQ_2\bigg)\bigg]- (\tilde{\cF})^{{a_0}ijkl} \bigg[ \phi _l p_3^{a_1} {p_3}_i (\veps_3^A)_{jk}\cQ\bigg]\bigg\}\labell{B3a}
\eeqa
The amplitude is   for $p=1$. We have explicitly calculated the amplitude \reef{amp23} for RR $(p+3)$-form and found exactly the above result. This amplitude is fully consistent with all Ward identities. As a result there is no amplitude for  RR $(p+5)$-form  which is also consistent with the S-matrix calculation.

The amplitudes \reef{Fp} and \reef{FF} contains the graviton and dilaton. For   graviton, the symmetric polarization tensor $(\varepsilon_3^S)_{\mu\nu}$ should be traceless, whereas, for dilaton it is given by   
 \beqa
(\varepsilon_3^S)_{\mu\nu}&=&\eta_{\mu\nu}-\ell_{\mu}(p_1)_{\nu}-\ell_{\nu}(p_1)_{\mu}
 \eeqa
  where the auxiliary field $\ell$ satisfies $\ell.p_1=1$ and should be canceled in the final amplitude. By replacing the  above polarization tensor in the amplitude, one finds the dilaton amplitude in the Einstein frame. We are interested, however,  in the dilaton amplitude in the string frame. To this end, we replace the graviton polarization in the amplitudes \reef{Fp} and \reef{FF} by $(\veps_3^S)_{\mu\nu}\rightarrow (\veps_3^S)_{\mu\nu}-\frac{1}{2}\eta_{\mu\nu}\Phi$ where $\Phi$ is the dilaton polarization which is one. The dilaton amplitudes resulting from this replacement should be added to the dilaton amplitude in the Einstein frame to produce the string frame amplitude for the dilaton.

\section{Contact terms at low energy }

The S-matrix elements that we have found in the previous  section, can be analyzed at low
energy to extract the appropriate couplings in field theory at order $\alpha'^2$. To this ends, one has to expand the integrals at low energy. The integrand of the integrals are invariant under  $SL(2,R)$ transformations. Fixing this symmetry, the explicit form of integrals  $\cQ$, $\cQ_1$ and $\cQ_2$ have been found in \cite{Garousi:2012gh} in terms of hypergeometric functions. Then the $\alpha'$ expansion produce the following expansions \cite{Garousi:2012gh,Garousi:2011ut}: 
\beqa
\cQ&=&\frac{2}{p_1.p_3}-\frac{\pi^2}{3}p_3.D.p_3+\cdots\nonumber\\
\cQ_1&=&-\frac{2}{3p_1.p_3}-\frac{8}{3p_1.D.p_1}+\frac{4}{3} \pi ^2 p_1.p_3+\frac{1}{2} \pi ^2 p_3.D.p_3-\frac{1}{6} \pi ^2 p_1.D.p_1 +\frac{16 \pi ^2 \left(p_2.p_3\right){}^2}{3 p_3.D.p_3} +\cdots
\nn\\
 \cQ_2&=&-\frac{2}{p_1.p_3}+\frac{\pi^2}{3}p_1.D.p_1+\cdots\labell{expand}
\eeqa
where dots refer to the terms with more than two momenta. They are related to the couplings at order $O(\alpha'^3)$ in which we are not interested. 
Similar calculations, produce the following expansion for the integrals \reef{int2}: 
\beqa
\cQ_3&=&\frac{2}{3p_1.p_3}+\frac{8}{3p_1.D.p_1}-\frac{4}{9} \pi ^2 p_1.p_3-\frac{1}{3} \pi ^2 p_3.D.p_3-\frac{2}{9} \pi ^2 p_1.D.p_1 +\frac{16 \pi ^2 \left(p_2.p_3\right){}^2}{9 p_1.D.p_1} +\cdots\nonumber\\
\cQ_4&=&\frac{4}{3p_1.p_3}+\frac{8}{3p_3.D.p_3}-\frac{4}{9} \pi ^2 p_1.p_3-\frac{2}{9} \pi ^2 p_1.D.p_1 +\frac{16 \pi ^2 \left(p_2.p_3\right){}^2}{9 p_1.D.p_1} +\cdots\labell{expand2}
\eeqa
The leading  massless poles in the open and closed string channels   should be reproduced by the supergravity couplings in the bulk and by the D-brane action  \reef{DBI} in which we are not interested in this paper.

The next to the leading order terms have  contact terms at order $\alpha'^2$, and massless poles   in the open string channel. It is consistent with the fact that the corrections to the type II supergravities   at order $\alpha'^2$ are zero. As a result, there is no massless closed string pole at order $\alpha'^2$.  The massless open string   poles should be reproduced by the D-brane action \reef{DBI} at order $\alpha'^0$, and by the couplings \reef{DBI2} and \reef{CS2f} at order $\alpha'^2$, \ie the Feynman amplitude is    
 \beqa
\cA =V_{RR}\,G_{NS}\,V_{NSNS}\labell{Feyn}
\label{feya}
\eeqa
where $V_{RR}$ is the vertex that includes RR form, $V_{NSNS}$ is the vertex that includes NSNS closed string and $G_{NS}$ is the open string propagator on the $D_p$-brane. One of these vertices should be calculated from \reef{DBI} and the other one should be calculated from \reef{DBI2} and \reef{CS2f}.  The Feynman diagram corresponding to the above amplitude is given figure \reef{feyd}. The standard forms of the gauge field and the transverse scalar propagators are 
\beqa
\label{propagator}
G_A^{ab}=\frac{-i\eta^{ab}}{(2\pi\alpha')^2 T_p \,p\inn V\inn p}&&G_{\phi}^{ij}=\frac{-i\eta^{ij}}{(2\pi\alpha')^2 T_p \,p\inn V\inn p}
\eeqa
 where $p$ is the open string momentum.
  
\begin{figure}[t!]
\centering
\includegraphics[scale=0.7]{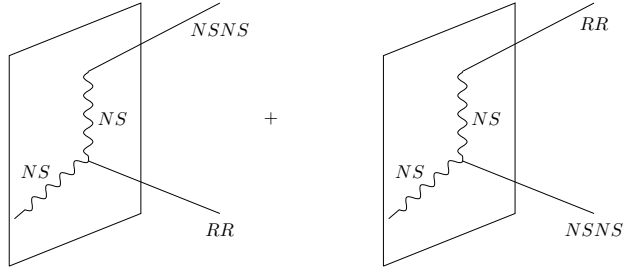}
\caption{One RR, one NSNS and one NS Feynman diagram}
\label{feyd}
\end{figure}

\subsection{RR ${(p-3)}$-form,  B-filed and  gauge field}

Using the above Feynman prescription, we have found the following amplitude between one RR $(p-3)$-form, one B-field and one open string gauge field: 
\beqa
\cA^{(p-3)} &=&-\frac{\pi^2\alpha'^2T_{p}}{24}\epsilon^{a_0a_1a_2a_3 a_4}\tilde{\cF}^{(2)}_{a_3a_4}\bigg(\frac{(p_2\inn p_3)^2}{p_3\inn V\inn p_3}\tilde{F}_{a_0a_1}p_3{}^{a}(\veps_3^A)_{aa_2}\label{cp-3f}\\
&&\hspace{11em}+\tilde{F}_{a_0a_1}(p_2\inn p_3)\, p_3{}^a(\veps_3^A)_{aa_2}\bigg)\nonumber
\eeqa
where  $\tilde{F}_{ab}=-ip_2{}^{[a}\veps_2{}^{b]}$ is the gauge field strength.
The first  term in the above Feynman amplitude is the open string massless pole which appears in the low energy limit of the string theory S-matrix element  \reef{B0a}.  The normalization of the above massless pole fixes the normalization of string amplitude \reef{B0a} to be   $\frac{i}{16}\alpha'^2T_p$. Note that we have chosen $p=4$ in above calculations.

The second term in  \reef{cp-3f} is a contact term which we call it $\cC^{(p-3)}_{BA}$. It   should   be subtracted from the contact terms of string theory amplitude at order $\alpha'^2$. There is another set of contact terms with structure of one RR $(p-3)$-form, one B-field and one gauge field in \reef{CS2f} that should be subtracted from the string theory contact terms. Subtracting these two sets of contact terms from  the string theory contact terms, we have found the following couplings for $p=4$:
\beqa
C^{(p-3)}_{BA} &=&-i\frac{\pi^2\alpha^2T_p}{12}\epsilon^{a_0a_1a_2a_3 a_4}\bigg[\frac{1}{2} {\tilde{\cF}}^{(2)}_{ba_4}(\veps_3^A)_{a_2a_3}(p_1\inn p_2)  p_1^{b}  p_2{}_{a_0} {\veps_2}_{a_1}
\labell{newp-3}\\\nn 
&&-\frac{1}{4} \tilde{\cF}^{(2)}_{ia_4}(\veps_3^A)_{a_2a_3}(p_1\inn V\inn p_1) p_2{}_{a_0} p_3^{i}{\veps_2}_{a_1}- \frac{1}{2}\tilde{\cF}^{(2)}_{ia_4}(\veps_3^A)_{a_2a_3} (p_1\inn p_2) p_2{}_{a_0} p_3^{i}{\veps_2}_{a_1}\\\nn
&&+\frac{1}{4}\tilde{\cF}^{(2)}_{a_3a_4}\bigg(2 (\veps_3^A)_{a_2i}  (p_1\inn V\inn p_1)  p_1^{i} p_2{}_{a_0} {\veps_2}_{a_1} +4 (\veps_3^A)_{a_2i}  (p_1\inn p_2)  p_1^{i}  p_2{}_{a_0} {\veps_2}_{a_1}\\\nn
&&- 4 (p_1\inn V\inn\veps_3^A)_{a_2}  (p_1\inn N \inn p_3)p_2{}_{a_0} {\veps_2}_{a_1} 
- 4 (p_2\inn\veps_3^A)_{a_2}  (p_1\inn N \inn p_3)  p_2{}_{a_0}  {\veps_2}_{a_1}\\\nn
&& + (\veps_3^A)_{a_1a_2}  (p_1\inn V\inn p_1) p_1\inn\veps_2 p_2{}_{a_0} -   (\veps_3^A)_{a_1a_2}(p_1\inn V\inn p_1) (p_1\inn  p_2) {\veps_2}_{a_0} \\\nn
&&+2 (p_2\inn\veps_3^A)_{ba_2}  (p_1\inn V\inn p_1)  p_2{}_{a_0} {\veps_2}_{a_1}\bigg)\bigg]
\eeqa
The above contact terms are new on-shell couplings in the momentum space at order $\alpha'^2$. 

\subsection{RR $(p-1)$-form, graviton/dilaton and  gauge field }

The Feynman amplitude of one RR $(p-1)$-form, one graviton and one gauge field produces exactly the massless poles of string theory amplitude \reef{Fp} at order $\alpha'^2$. It also produces some   contact terms. There are also contact terms of one RR $(p-1)$-form, one graviton and one gauge field in \reef{CS2f}.
Subtracting these    contact terms  from the string theory contact terms, we have found the following new contact terms at order $\alpha'^2$ in the string frame for $p=3$:
\beqa
C^{(p-1)}_{hA} &=&-\frac{\pi^2\alpha'^2T_{p}}{12}\epsilon^{a_0a_1 a_2a_3}\frac{1}{12} \bigg(3p_1{}_j \tilde{F}_{a_0a_1}\tilde{R}^{ij}\tilde{\cF}^{(3)}_{ia_2a_3}-2p_1{}_i\tilde{F}_{aa_0}\tilde{R}^{ai}\tilde{\cF}^{(3)}_{a_1a_2a_3}\labell{np-1hf}\\\nn
&&-6p_1{}_{a_0} \tilde{F}^{ab}\tilde{R}_{ba_1}\tilde{\cF}^{(3)}_{aa_2 a_3}+6p_1^b \tilde{F}_{aa_0} \tilde{R}^{ai}{}_{ba_1}\tilde{\cF}^{(3)}_{ia_2a_3}-2 p_1^i\tilde{F}^{ab}\tilde{R}_{aa_0bi} \tilde{\cF}^{(3)}_{a_1a_2a_3}\\\nn
&&+6p_1^a\tilde{F}_{a_0a_1} \tilde{R}_{aija_2}\tilde{\cF}^{(3)}_{ija_3}-6p_1{}_a\tilde{F}^{a}{}_{a_0}\tilde{R}^{i}{}_{a_1a_2}{}^{j}\tilde{\cF}^{(3)}_{ija_3}+3p_1^a \tilde{F}_{ab} \tilde{R}^{bi}{}_{a_0a_1}\tilde{\cF}^{(3)}_{ia_2 a_3}\\\nn
&&
-12p_1{}_{a_2}\tilde{F}^{a}{}_{a_0}\tilde{R}^{aij}{}_{a_1}\tilde{\cF}^{(3)}_{ija_3}\bigg)
\eeqa
where 
$\tilde{R}^{abcd}$ is the linearized Riemann curvature in the momentum space, \ie
\beqar
\tilde{R}^{abcd}&=&p_3^ap_3^c(\veps_3^S)^{bd}+p_3^bp_3^d(\veps_3^S)^{ac}-p_3^ap_3^d(\veps_3^S)^{bc}
-p_3^bp_3^c(\veps_3^S)^{ad}
\eeqar
    We have written the contact terms in terms of the linearized Riemann curvature to compact the form of contact terms.
	
Similar calculation for the dilaton, produces the following new contact terms in the string frame:
\beqa
C^{(p-1)}_{\Phi A} &=&\frac{\pi^2\alpha'^2T_{p}}{12}\epsilon^{a_0a_1a_2 a_3}\frac{1}{12} \bigg(3\tilde{\cF}^{(3)}_{ia_2a_3}p_3^ip_1\inn N\inn p_3 \tilde{F}_{a_0a_1}-2\tilde{\cF}^{(3)}_{a_1a_2 a_3}p_1\inn N\inn p_3p_3^a\tilde{F}_{aa_0}\label{np-1df}\\\nn
&&-6\tilde{\cF}^{(3)}_{aa_2a_3} p_3{}_bp_3{}_{a_1}p_1{}_{a_0}\tilde{F}^{ab}\bigg)
\eeqa
Note that the above dilaton contact terms are exactly reproduced by the graviton contact terms \reef{np-1hf} by replacing  $\tilde{R}^{ij}\rightarrow \tilde{R}^{ij}-p_3^ip_3^j$,  $\tilde{R}^{ai}\rightarrow \tilde{R}^{ai}-p_3^ap_3^i$ and  $\tilde{R}^{ab}\rightarrow \tilde{R}^{ab}-p_3^ap_3^b$, as expected from \reef{del}.

\subsection{RR $(p-1)$-form,  B-field and   scalar field }

The Feynman amplitude of one RR $(p-1)$-form, one B-field and one transverse scalar field produces exactly the corresponding massless poles of string theory amplitude \reef{Fp} at order $\alpha'^2$. It also produces the following   contact terms for $p=3$:
\beqa
\cC^{(p-1)}_{B\phi}&\!\!\!\!\!\!\!=\!\!\!\!\!\!\!&\frac{-i\pi^2\alpha^2T_{p}}{ 12}\epsilon^{a_0a_1a_2 a_3}\phi_i \bigg[\frac{1}{3!}\tilde{\cF}^{(3)}_{a_1a_2\cdots a_p}\bigg((p_1\inn V\inn\veps_3^A\inn p_2)p_3{}^ip_2{}_{a_0}
-2(p_2\inn\veps_3^A)_{a_0}(p_1\inn p_2)p_3{}^i\label{con4}\\\nonumber
&&-2(\veps_3^A)_{a_0i}(p_1\inn p_2)^2
+(p_2\inn\veps_3^A)_{i}(p_1\inn p_2) p_1{}_{a_0}-(p_2\inn\veps_3^A)_{i}(p_1\inn p_1) p_2{}_{a_0}
-3(p_2\inn\veps_3^A)_{i}(p_1\inn p_2) p_2{}_{a_0}\\\nonumber
&&+(p_2\inn\veps_3^A)_{i}(p_1\inn p_2) p_1{}_{a_0}\bigg)
-2\tilde{\cF}^{(3)}_{iaa_3}
(p_2\inn\veps_3^A)_{a_2}p_3{}^ap_1{}_{a_0}p_2{}_{a_1}\bigg]\labell{con2}
\eeqa
where $\phi^i$ is polarization of the transverse scalar fields. There are, however,  two other sets of couplings of one RR $(p-1)$-form, one B-field and one scalar field in the last line of action \reef{LTdual}. They are resulted from the projection operators  and the Taylor expansion operator implicit in \reef{LTdual}.

For the projection operators consider, for instance,  the coupling   $\prt^a\cF^{(p)}_{ia_2\cdots a_p}\prt^iH_{aa_0a_1}$. This coupling in terms of the projections of bulk tensors to the world volume and transverse spaces is 
\beqa
\bot^{\mu_1\nu_1}\prt_{a_0}X^{\sig_0}\cdots\prt_{a_p}X^{\sig_p}\prt_{a}X^{\rho_1}\prt^{a}X^{\rho_2}
\bigg(\prt_{\rho_1}\cF^{(p)}_{\mu_1 \sig_2\cdots \sig_p}\prt_{\nu_1}H_{\sig_0\sig_1}{}^{\rho_2}\bigg)\nn
\eeqa
where $\prt_aX^\mu$ is the projection operator into the world-volume space   and $\bot^{\mu\nu}$ is the projection operator into the transverse space, \ie  
\beqa
\bot^{\mu\nu}=G^{\mu\nu}-\tG^{\mu\nu}\,,\qquad\tG^{\mu\nu}=\frac{\prt X^{\mu}}{\prt{\sigma^a}}\frac{\prt X^{\nu}}{\prt{\sigma^b}}\tG^{ab}\,,
\eeqa
where $\tG^{\mu\nu}$  is the first fundamental form and 
$\widetilde{G}^{ab}$ is   inverse of the pull back metric.
In the static gauge, \ie $X^a=\sigma^a$ and $X^i=\phi^i$,   components of the projection operator $\bot^{\mu\nu}$ become  $\bot^{ab}=0$,  $\bot^{ai}=-\prt^a\phi^i$ and $\bot^{ij}=\eta^{ij}$ to the linear order of transverse scalar field in which we are interested. 

The closed string fields are function of spacetime coordinate $X^{\mu}$. In the static gauge, they    split into  world-volume
coordinates, $X^a=\sigma^a$, and transverse scalar fields $X^i=\phi^i$. Then the $\phi^i$ dependence of closed string fields appear in the world volume action via  Taylor expansion \cite{Garousi:1998fg}, \ie
$${\cal{C}_{\mu\nu\cdots}}(\phi^i)=\exp^{\Big[\phi^i\frac{\prt}{\prt x_i}\Big]}{\cal{C}}_{\mu\nu\cdots}^0(x^i)\Big|_{x^i=0}$$
where ${\cal{C}_{\mu\nu\cdots}}$ stands for any world volume or transverse derivative of a  massless closed string field.  

Using the projection operators, the couplings in the last line of \reef{LTdual} produce the following couplings of one RR $(p-1)$-form, one B-field and one scalar field: 
 \beqa
 S^{WS}_p &\!\!\!\!\!  \supset \!\!\!\!\!&-\frac{\pi^2\alpha^2T_{p}}{ 12}\frac{1}{2!\,(p!)}\int d^{p+1}x\,\eps^{a_0\cdots a_p}\Bigg(
\prt_{a_0}\phi^i\bigg(\prt^aH_{aij}\prt^j\cF^{(p)}_{a_1a_2\cdots a_p}-\prt^kH_{ijk}\prt^j\cF^{(p)}_{a_1a_2\cdots a_p}\nonumber\\
&&+p\,\prt^kH_{a_3jk}\prt^j\cF^{(p)}_{ia_1a_2\cdots a_p}
-p\,\prt^aH_{jaa_1}\prt^j\cF^{(3)}_{ia_2a_3\cdots a_p}
-p\,\prt^jH_{iaa_1}\prt^a\cF^{(p)}_{ja_2a_3\cdots a_p}\nonumber\\&&-p(p-1)\prt^jH_{aa_1a_2}\prt^a\cF^{(p)}_{ija_3\cdots a_p}\bigg)
-\prt_{a}\phi^i\bigg(\prt^bH_{iba_0}\prt^a\cF^{(p)}_{a_1a_2\cdots a_p}\nn\\
&&+\prt^bH^a{}_{ba_0}\prt_i\cF^{(p)}_{a_1a_2\cdots a_p}+\prt^jH_{jaa_0}\prt_i\cF^{(p)}_{a_1a_2\cdots a_p}-\prt^jH_{ija_0}\prt^a\cF^{(p)}_{a_1a_2\cdots a_p}
\nn\\
&&+p\,\prt_iH^a{}_{ba_0}\prt^b\cF^{(p)}_{a_1a_2\cdots a_p}+p\,\prt^aH_{ba_0a_1}\prt^b\cF^{(p)}_{ia_2a_3\cdots a_p}
-p\,\prt^jH^a{}_{a_0a_1}\prt_i\cF^{(p)}_{ja_2a_3\cdots a_p}
\nn\\
&&-p\,\prt^jH_{ia_0a_1}\prt^a\cF^{(p)}_{ja_2a_3\cdots a_p}
+2\prt_iH^{ja}{}_{a_0}\prt_j\cF^{(p)}_{a_1a_2\cdots a_p}
-2\prt^aH_{ija_0}\prt^j\cF^{(p)}_{a_1a_2\cdots a_p}\bigg)\Bigg)\labell{con8}
\eeqa
On the other hand,  the Taylor expansion produces the following couplings at the linear order of $\phi^i$:
\beqa
\label{con9}
S^{WS}_p &\!\!\!\!\!  \supset \!\!\!\!\!&-\frac{\pi^2\alpha^2T_{p}}{ 12}\frac{1}{2!\,(p-1)!}\int d^{p+1}x\,\eps^{a_0\cdots a_p}\phi^i\bigg(\prt_i\prt^a\cF^{(p)}_{ja_2\cdots a_p}\prt^jH_{aa_0a_1}+\prt^a\cF^{(p)}_{ja_2\cdots a_p}\prt_i\prt^jH_{aa_0a_1}\nn\\
&&-\prt_i\prt^j\cF^{(p)}_{a_1a_2\cdots a_p}(\prt^aH_{jaa_0}{}-\prt^kH_{a_0kj})-\prt^j\cF^{(p)}_{a_1a_2\cdots a_p}\prt_i(\prt^aH_{jaa_0}-\prt^kH_{jka_0})\bigg)\labell{taylor}
\eeqa
All these couplings should be produced by the corresponding contact terms of the amplitude \reef{Fp} at order $\alpha'^2$.

If the contact terms of string amplitude \reef{Fp} produce only the contact terms in \reef{con2} and the above couplings, then there would be no new couplings. However, we have subtracted the contact terms in \reef{con2}, \reef{con8} and \reef{taylor} from the corresponding contact terms of string amplitude \reef{Fp} at order $\alpha'^2$ and found the following extra contact terms for $p=3$:
\beqa
C^{(p-1)}_{B\phi}&\!\!\!\!\!\!\!=\!\!\!\!\!\!\!&-i\frac{\pi^2\alpha^2T_p}{12}\epsilon^{a_0a_1a_2 a_3}\phi^j\bigg[\tilde{\cF}^{(3)}_{ija_3} (p_2\inn\veps_3^A)_{a_2} p_1{}_{a_0} p_2{}_{a_1} p_3^{i} \labell{np-1bp}\\\nn
&&+\frac{1}{6}\tilde{\cF}^{(3)}_{a_1a_2 a_3}\bigg( (p_1\inn N\inn\veps_3^A)_{j} (p_1\inn p_2)p_2{}_{a_0}+(p_1\inn N\inn \veps_3^A\inn p_2 )p_2{}_{a_0} p_3^{j}+(p_1\inn N\inn p_3) ( p_2\inn\veps_3^A)_{j} p_2{}_{a_0} \bigg)
\\\nn
&&-\frac{1}{2} (\veps_3^A)_{a_1a_2}\tilde{\cF}^{(3)}_{ija_3\cdots a_p} (p_1\inn  p_2) p_2{}_{a_0} p_3^{i}+\frac{1}{6}\tilde{\cF}^{(3)}_{ja_2 a_3}\bigg(\frac{1}{2} (\veps_3^A)_{a_0a_1} (p_1\inn p_2)^2- (p_2\inn\veps_3^A)_{a_1}(p_1\inn p_2)  p_1{}_{a_0}\\\nn
&& -(p_1\inn V\inn \veps_3^A\inn p_2)p_1{}_{a_0}  p_2{}_{a_1}- (p_1\inn V\inn \veps_3^A)_{a_1} (p_1\inn p_2)  p_2{}_{a_0}+(p_2\inn \veps_3^A)_{a_1} (p_1\inn V\inn p_1)  p_2{}_{a_0}\bigg)\bigg]
\eeqa
 They indicates that there must be new couplings at order $\alpha'^2$. We will find them in section 4.

\subsection{RR $(p+1)$-form,  B-field and  gauge field }

The Feynman amplitude of one RR $(p+1)$-form, one B-field and one gauge field produces exactly the corresponding massless poles of string theory amplitude \reef{FF} at order $\alpha'^2$. It also produces the following   contact terms for $p=2$:
\beqa
\cC^{(p+1)}_{BA}=\frac{\pi^2\alpha^2T_{p}}{ 12}\frac{1}{3!}\epsilon^{a_0a_1a_2}\tilde{\cF}^{(4)}_{ia_0a_1a_2}\tilde{F}^{ab}\bigg[2(\veps_3^A)_{bi} p_1{}_a(p_2\inn p_3) 
 -(\veps_3^A)_{ab}(p_2\inn p_3) p_3{}^i\bigg]\nn
\eeqa
There is no contact term of one RR $(p+1)$-form, one B-field and one gauge field in \reef{CS2f}. Subtracting the above contact terms from the corresponding contact terms of \reef{FF} at order $\alpha'^2$, one finds the following on-shell couplings in the momentum space for $p=2$:
\beqa
C^{(p+1)}_{BA}&\!\!\!=\!\!\!&i\frac{\pi^2\alpha'^2T_{p}}{24}\frac{1}{ 4!} \epsilon^{a_0a_1a_2}\bigg[6p_3^kp_1{}_a\tilde{F}_{a_0a_1}\tilde{H}^{aij}\tilde{\cF}^{(4)}_{ijka_2}
+6p_3{}_bp_1{}_a\tilde{F}_{a_0}{}^{b}\tilde{H}^{aij}\tilde{\cF}^{(4)}_{ija_1 a_2}\labell{np+1bf}\\\nn
&&
-6p_3{}_ap_1{}_b\tilde{F}_{a_0}{}^{b}\tilde{H}^{aij}\tilde{\cF}^{(4)}_{ija_1 a_2}
+6p_1\inn V\inn p_3\tilde{F}_{a_0}{}^{b}\tilde{H}_{b}{}^{ij}\tilde{\cF}^{(4)}_{ija_1a_2}
+12p_3^kp_1{}_b\tilde{F}_{a_0}{}^{b}\tilde{H}_{a_1}{}^{ij}\tilde{\cF}^{(4)}_{ijka_2}\\\nn
&&
-12p_3^kp_1{}_{a_1}\tilde{F}_{a_0}{}^{b}\tilde{H}_{b}{}^{ij}\tilde{\cF}^{(4)}_{ijka_2}+4p_3{}_bp_1{}_c\tilde{F}^{ab}\tilde{H}_{a}{}^{ci}\tilde{\cF}^{(4)}_{ia_0a_1 a_2}
-4p_3{}_cp_1{}_b\tilde{F}^{ab}\tilde{H}_{a}{}^{ci}\tilde{\cF}^{(4)}_{ia_0a_1a_2}\\\nn
&&
-2p_1\inn V\inn p_3\tilde{F}^{ab}\tilde{H}_{ab}{}^{i}\tilde{\cF}^{(4)}_{ia_0a_1a_2}
-2p_3^jp_1{}_i\tilde{F}^{ab}\tilde{H}_{ab}{}^{i}\tilde{\cF}^{(4)}_{ja_0a_1 a_2}
-2p_1\inn N\inn p_3\tilde{F}^{ab}\tilde{H}_{ab}{}^{i}\tilde{\cF}^{(4)}_{ia_0a_1 a_2}
\\\nn
&&\hspace{11.4em}+6p_3{}_{a_0}p_1{}_b\tilde{F}^{ab}\tilde{H}_{a}{}^{ij}\tilde{\cF}^{(4)}_{ija_1  a_2}\bigg]
\eeqa
where $\tilde{H}^{\mu\nu\sig}=-2ip_3^{[\mu}(\veps_3^A){}^{\nu\sig]}$ is the field strength of B-field in the momentum space. Note that $H$ in the above contact terms carries  at least one transverse index. This is consistent with the observation that there is no couplings between one RR $(p+1)$ form and two gauge fields at order $\alpha'^2$ (see eq.\reef{CS2f}).

\subsection{RR $(p+1)$-form,  graviton/dilaton and  scalar field }

The Feynman amplitude of one RR $(p+1)$-form, one graviton/dilaton and one scalar field produces exactly the corresponding massless poles of string theory amplitude \reef{FF} at order $\alpha'^2$. It also produces the following   contact terms for graviton and dilaton, respectively:
\beqa
 \cC^{(p+1)}_{h\phi} &\!\!\!\!=\!\!\!\!&-\frac{\pi^2\alpha^2T_{p}}{12}\frac{i}{3!}\epsilon^{a_0a_1 a_2}\bigg\{\bigg[\tilde{\cF}^{(4)}_{ia_0a_1 a_2}2 (p_2\inn p_3)\bigg( \phi^{i}(p_1\inn V\inn\veps_3^S\inn p_2)-\phi^{i}(p_2\inn\veps_3^S\inn p_2)\labell{con6}\\
&&-p_3{}^{i}(\phi\inn N\inn\veps_3^S\inn p_2)+p_3\inn N\inn\phi
(p_2\inn\veps_3^S)^{i}+(p_2\inn p_3)(\phi\inn N\inn\veps_3^S)_{i}+(p_2\inn p_3)\, \phi^{i}\,\Tr(\veps_3^S\inn V)\bigg)
\nn\\ 
&&-(p_1\inn p_1)\phi^{i}(p_2\inn\veps_3^S\inn p_2)
\bigg]+3(p_2\inn p_3)p_2{}_{a_0}\phi^j\cF^{(4)}_{ija_1 a_2}\bigg[p_3{}^i\Tr((\veps_3^S)\inn V)-2(p_3\inn V\inn\veps_3^S)^{ai}\bigg]\bigg\}\nn\\
 \cC^{(p+1)}_{\Phi\phi}&\!\!\!\!=\!\!\!\!&-\frac{\pi^2\alpha^2T_{p}}{12}\epsilon^{a_0a_1 a_2}(p_2\inn p_3)\phi^i\bigg[\frac{i}{3!}\tilde{\cF}^{(4)}_{ia_0a_1 a_2} (p_2\inn p_3)+\frac{i}{2}\tilde{\cF}^{(4)}_{ija_1 a_2}p_2{}_{a_0}p_3{}^j\bigg]\nn
\eeqa
The projection operator and the Taylor expansion produce   two other sets of couplings of one RR $(p+1)$-form, one graviton/dilaton and one scalar field from the couplings in the second line of  action \reef{LTdual}. The projection operator produce the following couplings: 
\beqa
 S^{WS}_p &\!\!\!\!\!  \supset \!\!\!\!\!&\frac{\pi^2\alpha^2T_{p}}{ 12}\int d^{p+1}x\,\eps^{a_0\cdots a_p}\Bigg[\frac{\prt_{a_0}\phi^i}{p\,!}\bigg(p\,R_{a_1}{}^{ajk}\prt_a\cF^{(p+2)}_{ijka_2\cdots a_p}+2\cR^{jk}\prt_k\cF^{(p+2)}_{ija_1a_2\cdots a_p}\labell{ss1}\\\nonumber
&&R_{ai}{}^{jk}\prt^a\cF^{(p+2)}_{jka_1a_2\cdots a_p}\bigg)+\frac{2\,\prt_a\phi^i}{(p+1)!}\bigg(\cR_{i}{}^{j}\prt^a\cF^{(p+2)}_{ja_0a_1\cdots a_p}+R^{abj}{}_{i}\prt_b\cF^{(p+2)}_{ja_0a_1\cdots a_p}\\\nonumber
&&+\cR^{aj}\prt_i\cF^{(p+2)}_{ja_0a_1\cdots a_p}
+\cR^{a}{}_{j}\prt^j\cF^{(p+2)}_{ia_0a_1\cdots a_p}-R_{aji}{}^k\prt^j\cF^{(p+2)}_{ka_0a_1\cdots a_p}-R_{ika}{}^{j}\prt^k\cF^{(p+2)}_{ja_0a_1\cdots a_p}\\\nonumber
&&-\frac{(p+1)}{2}R_{a_0i}{}^{jk}\prt^a\cF^{(p+2)}_{jka_1a_2\cdots a_p}
+R^{aj}{}_{a_0b}\prt^b\cF^{(p+2)}_{ija_1a_2\cdots a_p}+\frac{(p+1)}{2}R^{jka}{}_{a_0}\prt_i\cF^{(p+2)}_{jka_1a_2\cdots a_p}\bigg)\Bigg]\label{con10}
\eeqa
The Taylor expansion produces the following couplings:
\beqa
S^{WS}_p &\!\!\!\!\!  \supset \!\!\!\!\!&-\frac{\pi^2\alpha^2T_{p}}{ 12}\frac{1}{\,p\,!}\int d^{p+1}x\,\eps^{a_0\cdots a_p}\phi^k\bigg(\frac{1}{2!}\prt_k\prt_a\cF^{(p+2)}_{ija_1\cdots a_p}R_{a_0}{}^{aij}\labell{con11}\\\nonumber
&&+\frac{1}{2!}\prt_a\cF^{(p+2)}_{ija_1\cdots a_p}\prt_k R_{a_0}{}^{aij}
+\frac{2}{p+1}\prt_k\prt_j\cF^{(p+2)}_{a_0\cdots a_pi}{\cR}^{ij}+\frac{2}{p+1}\prt_j\cF^{(p+2)}_{a_0\cdots a_pi}\prt_k{\cR}^{ij}\bigg)
\eeqa
There is still another set of couplings in the second line of \reef{LTdual} which are resulted from the $\Omega\Omega$ terms in $R_N$ and $\bar{\cR} $. If one considers the linear graviton in one of $\Omega$ and the transverse scalar field in the another $\Omega$, then one finds couplings of one RR $(p+1)$-form, one graviton and one transverse scalar field. All these couplings should be produced by the corresponding contact terms of the amplitude \reef{FF} at order $\alpha'^2$.

We have subtracted all above  field theory contact terms   from the corresponding string theory contact terms in \reef{FF}. We have found that the field theory contact terms for dilaton are exactly the same as the string theory contact terms. However, for graviton we have found the following extra contact terms in the string frame for $p=2$:
\beqa
C^{(p+1)}_{h\phi}&\!\!\!=\!\!\!&i\frac{\pi^2\alpha'^2T_{p}}{12} \epsilon^{a_0a_1 a_2}\tilde{\cF}^{(4)}_{ijka_2}\tilde{R}^{bjk}{}_{a_1}\tilde{\Om}_{ba_0}{}^{i}\labell{cp+1hp0} 
\eeqa
Our result that there is no extra couplings for dilaton is consistent with the above couplings for graviton because the dilaton contact terms should be produced by the graviton contact terms in which one replaces the Ricci curvature with the second derivative of dilaton. There is no Ricci curvature in the above contact term. As a result there should be no contact term for dilaton.

\subsection{RR ${(p+3)}$-form, B-field and  scalar field}

The sting theory amplitude \reef{B3a} has no massless pole at order $\alpha'^2$. This can be seen from the $\alpha'$-expansion of $\cQ$'s in \reef{expand}. The integrals $\cQ$ and $\cQ_2$ have no massless pole and the massless pole in $\cQ_1$ produces contact term after multiplying it with $p_3\inn V\inn p_3$ which appear in the string amplitude \reef{B3a}. Therefore, the amplitude \reef{B3a} produces only contact terms at order $\alpha'^2$. This is consistent with the fact that the Feynman amplitude \reef{Feyn} is zero in this case. There is no coupling in \reef{DBI} or \reef{CS2f} which contains the   RR $(p+3)$-form.  

However, the projection operator and the Taylor expansion produce   the following   couplings of one RR $(p+3)$-form, one B-field and one scalar field from the coupling in the first line of  action \reef{LTdual}: 
\beqa
S^{WS}_p &\!\!\!\!\!  \supset \!\!\!\!\!&-\frac{\pi^2\alpha^2T_{p}}{ 12}\frac{1}{3!(p+1)!}\int d^{p+1}x\,\eps^{a_0\cdots a_p}\Bigg(\phi^i\bigg[\prt_i\prt_a\cF^{(p+4)}_{ia_0\cdots a_pjk}\prt^aH^{ijk}\labell{con13}\\\nonumber
&&+\prt_aF^{(p+4)}_{ia_0\cdots a_pjk}\prt_i\prt^aH^{ijk}\bigg]
+\prt_a\phi^i\bigg[\prt_iH^{jkl}\prt^a\cF^{(p+4)}_{jkla_0a_1\cdots a_p}+\prt^aH^{jkl}\prt_i\cF^{(p+4)}_{jkla_0a_1\cdots a_p}\\\nonumber
&&3\,\prt_bH^{jka}\prt^b\cF^{(p+2)}_{ijka_0a_1\cdots a_p}\bigg]
-(p+1)\prt_{a_0}\phi^i\prt_a\cF^{(5)}_{ijkla_1}\prt^aH^{jkl}\Bigg)
\eeqa
Subtracting the above couplings from the contact terms of \reef{B3a}, we have found the following new on-shell couplings in the momentum space for $p=1$: 
\beqa
C^{(p+3)}_{B\phi}&\!\!\!=\!\!\!&\frac{\pi^2\alpha'^2T_{p}}{12} \epsilon^{a_0a_1}\frac{i}{4}\bigg[
2\tilde{\cF}^{(5)}_{ijkla_1}p_2\inn p_3  p_2{}_{a_0}  p_3^{i}(\veps_3^A)^{kl} \phi^{j}\labell{np+3bf}\\\nn
&&+(p_2\inn p_3 )^2\tilde{\cF}^{(5)}_{ijka_0a_1}(\veps_3^A)^{jk}\phi^{i}+2p_2\inn p_3\tilde{\cF}^{(5)}_{ijka_0a_1} p_2^{b} p_3^{i}(\veps_3^A)_{b}{}^{k}\phi^{j}\bigg]
\eeqa
The string amplitude \reef{B3a} has no gravity or dilaton couplings, so there is no contact terms for gravity or dilaton.

\section{New couplings}

In this section we are interested in the effective action $S_p$ which are linearly covariant under general coordinate transformations, invariant under linear T-duality and are consistent with the contact terms of the S-matrix elements that we have found in the previous section. It has been   argued in \cite{ Robbins:2014ara} that to construct the effective action   for probe D$_p$-branes, one has to impose   the bulk equations of motion  at order $\alpha'^0$.  Since we are interested in the world volume couplings which have linear closed string fields, we have to impose the supergravity equations of motion at the linear order. In the string frame they are
\beqa
R+4\nabla^2\Phi&=&0\nonumber\\
R_{\mu\nu}+2\nabla_{\mu\nu}\Phi&=&0\nonumber\\
\nabla^{\rho}H_{\rho\mu\nu}&=&0\nonumber\\
\nabla^{\mu_1}\cF^{(n)}_{\mu_1\mu_2\cdots \mu_n}&=&0
\eeqa
where $\mu,\nu,\rho$ are the bulk indices.  Using these equations, one may rewrite the couplings in which two
normal indices are contracted within a single field (including the derivatives acting on that field) in terms of couplings in which two world volume indices contracted, \ie 
\beqa
R^i{}_{\mu i\nu }  &=&  -2\nabla_{\mu\nu}\Phi-R^a{}_{\mu a\nu }\nonumber\\
\nabla^i{}_i\Phi&=&-\nabla^a{}_a\Phi\nonumber\\
\nabla^{i}H_{i\mu\nu}&=&-\nabla^{a}H_{a\mu\nu}\nonumber\\
\nabla^{i}\cF^{(n)}_{i\mu_2\cdots \mu_n}&=&-\nabla^{a}\cF^{(n)}_{a\mu_2\cdots \mu_n}\labell{eom}
\eeqa
This indicates that the terms on the left-hand side are not    independent. In other words, the coefficients of the couplings in $S_p$ which involve the terms on the left-hand side are not independent. It turns out that if one considers the terms on the left-hand side as independent, then the  effective action would not be invariant under linear T-duality.  
 
We are interested in the new couplings of two closed strings and one open string at order $\alpha'^2$. So each term must have five derivatives. Using the fact that each field must have at least one derivative, one concludes that the maximum number of derivatives on a field must be three. On the other hand, no contact terms in the previous section has  three momentum in the transverse space. So   at least one of the three derivatives must be a world volume derivative. Using integration by part one may  rewrite it  in terms of two derivatives. So up to total derivative terms, we have to consider all contractions of two closed and one open strings in which each field has at most two derivatives.

\subsection{All contractions of one RR, one NSNS and one NS fields}
 
In this subsection,   using the Mathematica package "xAct" \cite{Nutma:2013zea}, we write all contractions of one massless RR, one NSNS and one NS fields in which each field has at most two derivatives. We consider the structures that are produced by the   S-matrix elements at order $\alpha'^2$ in the previous section.

The couplings of one 
  RR ${(p-3)}$-form, one B-field and one gauge field have three structures, \ie $ \cF^{(p-2)}\prt H \prt F$, $\prt\cF^{(p-2)}\prt H F$ and $\prt\cF^{(p-2)} H\prt F$. Each structure has the following contractions: 
  \beqa
\label{cp-3bf}
\cL^{(p-3)}_{ 1BF}&\sim&\epsilon^{a_0a_1a_2a_3 a_4}\bigg(\delta_1\cF^{(2)}_{ab}\prt^a F_{a_0a_1}\prt^b H_{a_2a_3a_4}+\delta_2
{\cF}^{(2)}_{aa_0}\prt^b F_{ba_1}\prt^a H_{a_2a_3a_4}+\cdots\\\nonumber
&&\hspace{14.1em}+\delta_{46}{\cF}^{(2)}_{ab}\prt^a F_{a_0a_1}\prt^b H_{a_2a_3a_4}\bigg)\\\nonumber
\cL^{(p-3)}_{ 2BF}&\sim&\epsilon^{a_0a_1a_2a_3 a_4}\bigg(\nu_1\prt^b\cF^{(2)}_{a_3 a_4}\prt^a F_{ab} H_{a_0a_1a_2}+\nu_2
\prt^b\cF^{(2)}_{ba_4}\prt^a F_{a_0a} H_{a_1a_2a_3}+\cdots\\\nonumber
&&\hspace{14.4em}+\nu_{47}\prt^i\cF^{(2)}_{a_3 a_4}\prt^a F_{a_0a_1} H_{iaa_2}\bigg)\\\nonumber
\cL^{(p-3)}_{3BF}&\sim&\epsilon^{a_0a_1a_2a_3 a_4}\bigg(\mu_1\prt^a\cF^{(2)}_{a_3 a_4}F_{ab} \prt^bH_{a_0a_1a_2}+\mu_2
\prt^b\cF^{(2)}_{ba_4} F_{a_0a} \prt^aH_{a_1a_2a_3}+\cdots\\\nonumber
&&\hspace{14.4em}+\mu_{75}\prt^j\cF^{(2)}_{ia_4} F_{a_0a_1} \prt_jH_{ia_2a_3}\bigg)
\eeqa
where $\delta_i$, $\nu_i$ and $\mu_i$ are 168 arbitrary coefficients that should be determined by imposing appropriate  constraints. Imposing the Bianchi identities and ignoring total derivative terms, one finds that all these coefficients are not independent. One may   first find independent coefficients and then impose the constraints. Or one may first impose the constraints and then ignore the terms that are related by the Bianchi identities and total derivative terms. We use the latter approach which is easier to work with  computer. Note that we have used the B-field only in the form of field strength $H$. The B-field also appears in the form of $B_{ab}$ or $\prt_aB_{bc}$. However, the coefficients of such  couplings have been  already found to be  \reef{CS2f} by studying the S-matrix element of one closed and two open strings \cite{Jalali:2015xca}.    

The couplings of  one RR $(p-1)$-form, one graviton and one gauge field which have at most two derivatives on each field, have two structures   $\prt\cF^{(p)}R F$ and  $\cF^{(p)}R \prt F$. All contractions of these structures are the following: 
\beqa
\label{cp-1hf}
\cL^{(p-1)}_{ 1hF}&\sim&\epsilon^{a_0a_1a_2 a_3}\bigg(\zeta_1\cF^{(3)}_{ca_0a_1}R^{bc}{}_{a_2a_3}\prt^a F_{ab}-\zeta_2
{\cF}^{(3)}_{ia_0a_1}R_{ba_2a_3i}\prt_a F^{ab}+\cdots\\\nonumber
&&\hspace{14.1em}+\zeta_{99}{\cF}^{(3)}_{aa_2a_3}\cR_{bc}{}^{bc} \prt^a F_{a_2a_3}\bigg)\\\nonumber
\cL^{(p-1)}_{2hF}&\sim&\epsilon^{a_0a_1a_2 a_3}\bigg(\rho_1\prt^a\cF^{(3)}_{ca_2a_3}R^{bc}{}_{a_0a_1} F_{ab}+\rho_2
\prt_a\cF^{(3)}_{ca_2 a_3}R_{ba_0}{}^c{}_{a_1}F^{ab}+\cdots\\\nonumber
&&\hspace{13.8em}+\rho_{157}\prt_i\cF^{(3)}_{jka_3}R^{a_2ijk} F_{a_0a_1} \bigg)
\eeqa
where  $\rho_i$ and $\z_i$   are 256 unknown coefficients which should be fixed by appropriate constraints. Note that we have used the graviton only in the form of curvature $R$. The graviton also appears in the form of second fundamental form. However, the coefficients of such  couplings have been  already found to be  \reef{CS2f} by studying the S-matrix element of one closed and two open strings \cite{Jalali:2015xca}. 

The couplings of  one RR $(p-1)$-form, one H-field and one scalar field have two structures   ${\cF}^{(p)}\prt H \Omega$ and  $\prt\cF^{(p)}\prt H \Omega$. All contractions of these structures are the following:
  \beqa
\label{cp-1Bp}
\cL^{(p-1)}_{ 1B\phi}&\sim&\epsilon^{a_0a_1a_2 a_3}\bigg(\tau_1\cF^{(3)}_{iba_3}\Omega_a{}^{ai}\prt^b H_{a_0a_1a_2}+\tau_2
{\cF}^{(3)}_{ba_2a_3}\Omega_a{}^{ai}\prt^b H^{a_0a_1i}+\cdots\\\nonumber
&&\hspace{14.4em}+\tau_{49}{\cF}^{(3)}_{jaa_3}\Omega_{a_0}{}^{ai} \prt^j H_{a_1a_2i}\bigg)\\\nonumber
\cL^{(p-1)}_{ 2B\phi}&\sim&\epsilon^{a_0a_1a_2 a_3}\bigg(\la_1\prt^b\cF^{(3)}_{a_1a_2a_3}\Omega_a{}^{ai} H_{ba_0i}+\la_2
\prt^b\cF^{(3)}_{ia_2a_3}\Omega_a{}^{ai}H^{ba_0a_1}+\cdots\\\nonumber
&&\hspace{13.8em}+\la_{49}\prt^j\cF^{(3)}_{ia_2a_3}\Omega_{a_0}{}^{ai} H_{aa_1j} \bigg)
\eeqa
where $\tau_i$ and $\la_i$ are the unknown coefficients and  $\Omega$ is the second fundamental form   in the static gauge, \ie $\Omega_{ab}^i=\prt_a\prt_b\phi^i-\Gamma_{ab}{}^c\prt_c\phi^i+\Gamma_{ab}{}^i$, and we have  considered  only the linear   scalar   part of it. Note that we have used the   scalar fields only in the form of the second fundamental form. The scalar fields also appear in the form of Taylor expansion and pull-back operator of two closed strings.  However, the coefficients of such  couplings have been  already found to be  \reef{LTdual} by studying the S-matrix element of two closed   strings \cite{Garousi:2010ki}. 

All contractions of one RR$(p-1)$-form, one dilaton and one gauge field are the following: 
\beqa
\label{cp-1df}
\cL^{(p-1)}_{ 1\Phi F}&\sim&\epsilon^{a_0a_1a_2 a_3}\bigg(\pi_1\cF^{(3)}_{a_1a_2a_3}\prt_a\prt^a\Phi \prt^bF_{ba_0}+\pi_2
{\cF}^{(3)}_{ba_2a_3}\prt^a\prt^b\Phi \prt_a F^{a_0a_1}+\cdots\nonumber\\\nonumber
&&\hspace{13.8em}+\pi_{21}\cF^{(3)}_{iba_3}\prt^i\prt_{a_0}\Phi\prt^b F_{a_1a_2} \bigg)\\\nonumber
\cL^{(p-1)}_{ 2\Phi F}&\sim&\epsilon^{a_0a_1a_2 a_3}\bigg(\vartheta_1\prt^a\cF^{(3)}_{a_1a_2a_3}F_{aa_0}\prt^b\prt_b\Phi  +\vartheta_2
\prt^a\cF^{(3)}_{a_1a_2a_3}F_{ab}\prt^b\prt_{a_0}\Phi  +\cdots\\\nonumber
&&\hspace{14.1em}+\vartheta_{30}\prt_j\cF^{(3)}_{ia_2a_3}F_{a_0a_1}\prt^i\prt^j\Phi \bigg)
\\\nonumber
\cL^{(p-1)}_{ 3\Phi F}&\sim&\epsilon^{a_0a_1a_2 a_3}\bigg(\varsigma_1\prt^a\cF^{(3)}_{a_1a_2a_3}\prt^bF_{ba_0}\prt_a\Phi  +\varsigma_2
\prt^b\cF^{(3)}_{a_1a_2a_3}\prt^aF_{ba_0}\prt_a\Phi +\cdots\\\nonumber
&&\hspace{14.1em}+\varsigma_{43}\prt^i\cF^{(3)}_{a_1a_2a_3}\prt^aF_{aa_0}\prt_i\Phi  \bigg)\\
\eeqa
where $\pi_i$, $\vartheta_i$ and $\varsigma_i$ are the coefficients and we have used the observation that each field should appear with  one or two derivatives.

All contractions of one RR ${(p+1)}$-form, one B-field and one gauge field are  the following:
\beqa
\label{cp+1Bf}
\cL^{(1)}_{C^{(p+1)}Bf}&\sim&\epsilon^{a_0a_1a_2}\bigg(\ga_1\cF^{(4)}_{ica_1a_2}\prt^bH^{ic}{}_{a_0}\prt^aF_{ab}+\ga_2
{\cF}^{(4)}_{ija_1a_2}\prt^bH_{a_0}{}^{ij}\prt^aF_{ab}+\cdots\\\nonumber
&&\hspace{13.5em}+\ga_{228}\tilde{\cF}^{(4)}_{ijkl}\prt^kH^{ijk}\prt_{a_2}F_{a_0a_1} \bigg)\\\nonumber
\cL^{(2)}_{C{(p+1)}Bf}&\sim&\epsilon^{a_0a_1a_2}\bigg(\iota_1\prt^b\cF^{(4)}_{ica_1 a_2} H^{ic}{}^{a_0}\prt^aF_{ab}+\iota_2
\prt^b\cF^{(4)}_{ija_1 a_2}H^{ij}{}_{a_0}\prt^aF_{ab}+\cdots\\\nonumber
&&\hspace{14.1em}+\iota_{222}\prt_k\cF^{(4)}_{ijca_2}H^{ijk}\prt^cF_{a_0a_1} \\\nonumber
\cL^{(2)}_{C{(p+1)}Bf}&\sim&\epsilon^{a_0a_1a_2}\bigg(\ka_1\prt^b\cF^{(4)}_{ica_1 a_2} \prt^aH^{ic}{}^{a_0}F_{ab}+\ka_2
\prt^b\cF^{(4)}_{ija_1 a_2}\prt^aH^{ij}{}_{a_0}F_{ab}+\cdots\\\nonumber
&&\hspace{14.1em}+\ka_{349}\prt_l\cF^{(4)}_{ijka_2}\prt^lH^{ijk}F_{a_0a_1} \bigg)
\eeqa
In this case we have 799 coefficients. Note that all indices of the RR field strength can not be world volume indices because two indices become identical which make the antisymmetric field strength to be zero. As a result at least one of the indices of this tensor must be transverse index. This index must be contracted with B-field or its derivative. Therefore, the B-field can not be in the form of $B_{ab}$, and its world volume derivatives. This is consistent with the fact that there is no such couplings in \reef{CS2f}.

All contractions of one  RR $(p+1)$-form, one graviton and one scalar field are the following:
\beqa
\cL^{(p+1)}_{1h\phi}&\sim&\epsilon^{a_0a_1a_2}\bigg(\psi_1\cF^{(4)}_{ibca_2 }R_{a_0a_1}{}^{bc}\Omega_a{}^{ai}+2\psi_2
{\cF}^{(4)}_{ijba_2}R_{a_0a_1}{}^{jb}\Omega_a{}^{ai}+\cdots\labell{cp+1hp}\nonumber\\
&&\hspace{13em}+\psi_{82}{\cF}^{(4)}_{jba_1 a_2}R_{ai}{}^{aj}\Omega_{a_0}{}^{bi} \bigg)
\eeqa
where $\psi_i$ are the unknown coefficients. Note that the graviton can also appear in the second fundamental form which produces couplings with structure $\prt\cF\Omega\Omega$. Such couplings, however,  are appeared in \reef{LTdual} which have been found in \cite{Jalali:2015xca} by studying the S-matrix element of one closed and two open strings. Similarly, the scalar fields can appear as pull-back and Taylor expansion of two closed string couplings which have been already considered in \reef{LTdual}.

All contractions of  one RR $(p+1)$-form, one dilaton and one scalar field are the following:
\beqa
\cL^{(p+1)}_{1\Phi\phi}&\sim&\epsilon^{a_0a_1a_2}\bigg(\Pi_1\cF^{(4)}_{ia_0a_1 a_2}\Om_b{}^{bi}\prt_a\prt^a\Phi +\Pi_2
{\cF}^{(4)}_{ia_0a_1 a_2}\Om_{ab}{}^i\prt^a\prt^b\Phi +\cdots\labell{cp+1dp}\\\nonumber
&&\hspace{13.6em}+\Pi_{17}\cF^{(4)}_{jba_1a_2}\Om_{a_0}{}^{bi}\prt^j\prt_{i}\Phi \bigg)\\
\cL^{(p+1)}_{2\Phi\phi}&\sim&\epsilon^{a_0a_1a_2}\bigg(\Lambda_1\prt_a\cF^{(4)}_{ia_0a_1 a_2}\Om_b{}^{bi}\prt^a\Phi +\Lambda_2\prt^a\cF^{(4)}_{ia_0a_1 a_2}\Om_{ab}{}^{i}\prt^b\Phi +\cdots\nn\\\nonumber
&&\hspace{13.5em}+\Lambda_{24}\prt_j\cF^{(4)}_{iba_1 a_2}\Om_{ba_0}{}^{j}\prt^i\Phi  \bigg)
\eeqa
where $\Pi_i$ and $\Lambda_i$ are the unknown coefficients. Here again the scalar fields appear in the second fundamental form. The presence of these fields in the Taylor expansion and pull-back of \reef{LTdual} have been already considered.

Finally, all contractions  of one RR $(p+3)$-from, one B-field and one scalar field are the following:
\beqa
\cL^{(p+3)}_{1B\phi}&\sim&\epsilon^{a_0a_1 }\bigg(\theta_1\cF^{(5)}_{ijkba_1 }\prt_{a_0}H^{bjk}\Omega_a{}^{ai}+\theta_2
{\cF}^{(5)}_{ijkla_1}\prt_{a_0}H^{jkl}\Omega_a{}^{ai}+\cdots\labell{cp+3bp}\\\nonumber
&&\hspace{13.5em}+\theta_{89}{\cF}^{(5)}_{jklba_1}\prt^lH^{ijk}\Omega_{a_0}{}^{bi} \bigg)\\\nonumber
\cL^{(p+3)}_{2B\phi}&\sim&\epsilon^{a_0a_1}\bigg(\omega_1\prt_{a_0}{\cF}^{(5)}_{ijkba_1} H^{bjk}\Omega_{a}{}^{ai}+\omega_2
\prt_{a_0}{\cF}^{(5)}_{ijkla_1}H^{jkl}\Omega_a{}^{ai}+\cdots\\\nonumber
&&\hspace{14.1em}+\omega_{85}\prt_l\cF^{(5)}_{ijkba_1}H^{jkl}
\Omega_{a_0}{}^{bi}\bigg)
\eeqa
where $\theta_i$ and $\omega_i$ are the unknown coefficients and  the scalar fields appear in the second fundamental form. Since the scalar fields appear in the above couplings through the second fundamental form, the derivative of the second fundamental form has three world volume derivatives which can be converted to two derivatives by using integration by part. As a result, up to total derivative terms, the couplings with structure $\cF H D\Omega$ are not independent of the above couplings.

\subsection{S-matrix constraint}

We have found all contractions of one RR, one NSNS and one NS fields in the previous section. Their coefficients should be found by imposing appropriate constraints. One constraint is the fact that when imposing the on-shell relations on the couplings, they must be identical to the contact terms that we have found in section 3. 

The   couplings in \reef{cp-3bf} must be identical to the contact terms in \reef{newp-3} after using the on-shell relations. This produces the following relations between the coefficients:  
\beqa
&&\mu_{65}=-{\alpha}/{4}-2 \mu_{10}-\mu_{14}+\mu_{27}-2 \mu_{62}+2\mu_{63},\ \mu_{67}={\alpha}/{8}-\mu_{10}-\mu_{51}-\mu_{62}\labell{cons1}\\
&&\nu_{29}={\alpha}/{8}+3\delta_1+\delta_{10}-\delta_{12}/2+\delta_{30}-3\delta_5/2+\mu_{11}-\mu_{12}-\mu_{13}/2-\mu_{15}/2
+\mu_{16}
-\mu_{17}/2\nn\\
&&-2\mu_2/3+\mu_{20}-3\mu_3/2-\mu_{33}/2-\mu_{38}/2+\mu_{39}/2-3\mu_4
-\mu_{40}+3\mu_6/2+3\nu_{13}/2-3\nu_{14}\nonumber\\
&&+\nu_{25}/2+\nu_{26}/2-\nu_{28}\,,
 \nu_{37}
={\alpha}/{8}+\delta_{13}/2-\delta_{14}-\delta_{20}/2+\delta_{21}+\delta_{27}/4-\delta_{28}/2-\mu_{12}
+\mu_{17}/2\nonumber\\
&&-\mu_{20}+\mu_{23}/2-\mu_{24}-\mu_{31}/4+\mu_{32}/2
-\nu_{15}/2+\nu_{16}-\nu_{34}/4+\nu_{35}/2+\nu_{36}/2\nn\\
&&
\mu_9={\alpha}/{8}-\mu_{10}+\mu_{29}+\mu_{49}-\mu_{51}+3\mu_{53}\nonumber\\
 && \nu_9={\alpha}/{8}+2\delta_{10}/3-\delta_{12}/3-\delta_3-\delta_{34}/3-\delta_{37}/3-\delta_7
-2\mu_{12}/3-\mu_{2}-\mu_{33}/3\nn\\
&&-\nu_{15}/3+2\nu_{16}/3-\nu_{3}\,,
\mu_{7}={\alpha}/{12}+2\mu_{11}/3-\mu_{13}/3-\mu_{15}/3+2\mu_{16}/3
-\mu_{17}/3-\mu_2\nonumber\\
&&+2\mu_{20}/3-\mu_3-\mu_{33}/3-\mu_{38}/3-\mu_{41}/3
\nonumber\\
&& \nu_5={\alpha}/{8}+\delta_{10}-\delta_{12}/2+\delta_{16}+\delta_{19}+\mu_{11}-\mu_{12}-\mu_{13}/2-\nu_{10}-\nu_{15}/2
+\nu_{16}+\nu_{24}/2+\nu_{27}/2\nn\\
&&
\nu_6={\alpha}/{2}+6\delta_1+\delta_{13}-2\delta_{14}+\delta_{15}-2\delta_{18}+2\delta_{30}-\delta_{31}-3\delta_{5}-2\mu_{12}
-3\mu_2\nonumber\\
&&-3\mu_3-\mu_{33}-\mu_{38}+\mu_{39}
-6\mu_4-2\mu_{40}+3\mu_6+2\nu_{11}+3\nu_{13}-6\nu_{14}-\nu_{15}+2\nu_{16}
+\nu_{25}-2\nu_{28}\nonumber\\
&&\nu_7={\alpha}/{4}
+2\delta_{24}-\delta_{26}+\delta_{41}-2\delta_{43}+6\delta_{44}-3\delta_{46}+\delta_8-2\delta_9-\mu_{28}-2\mu_{29}
+\mu_{30}-3\mu_{52}\nonumber\\
&&-6\mu_{53}+3\mu_{54} +2\nu_{12}+\nu_{19}-2\nu_{21}
,\  \nu_{47}={\alpha}/{4}+\mu_{14}/2
-\mu_{25}/4+\mu_{48}/2-\mu_{63}+\mu_{64}/2\nonumber\\
&&-\mu_{66}-\nu_{32}/4
+\nu_{33}/2+\nu_{46}/2\,,\mu_{21}=-\mu_{12}-\mu_{18},\  \mu_{75}=- \mu_{74},\ \nu_{44}=\delta_{24}-\delta_{26}/2+3\delta_{44}\nonumber\\
&&-3\delta_{46}/2-\mu_{29}+\mu_{30}/2-3\mu_{53}
 +3\mu_{54}/2+\mu_{59}/2+3\mu_{60}/2
+\nu_{19}/2-\nu_{21}+\nu_{43}/2\nn
\eeqa
where $\alpha=-\frac{T_p\pi^2\alpha'^2}{12}$. 
 
The   couplings in \reef{cp-1hf} must be identical to the contact terms in \reef{np-1hf} after using the on-shell relations. This produces the following relations between the coefficients: 
\beqa
 && \rho_{19} =\alpha/2+\zeta_{33}-\zeta_{48}/2-2\zeta_{65}+\zeta_{74}+\rho_{146} +\rho_{147}/2 - 2 \rho_{18},\ \rho_{ 56 }=\alpha/2 -\rho_{ 151} - 2 \rho_{ 152} - 2 \rho_{ 55}\nn\\\nonumber
 && \rho_{80}=-\alpha/2+\rho_{133}-3\rho_{139}+\rho_{16},\  \rho_{ 52} =\alpha+2\zeta_{33}-\zeta_{48}-4\zeta_{65}+2\zeta_{74} - 2 \rho_{ 148} + 2 \rho_{ 149} + 2 \rho_{ 51}
\\\nn
&&\rho_{44}=\alpha/2+2\zeta_{23}+\zeta_{26}-\zeta_{31}/2+\zeta_{32}/2+\zeta_{35}+\zeta_{45}+\zeta_{46}/2
+\zeta_{47}-\zeta_{55}/2-\zeta_{59}+\zeta_{63}-\zeta_{72}\\\nn
&&-\rho_{12}-2\rho_{13}+2\rho_{144}
-\rho_{43}/2,\ \rho_{130}=\alpha/4+\zeta_{23}+\zeta_{26}/2+\zeta_{35}/2+\zeta_{45}/2-\rho_{12}/2-\rho_{129}/2\\\nonumber
&&-\rho_{13}\,,
\rho_{77}=-\alpha+2\zeta_{23}+\zeta_{26}+\zeta_{35}+\zeta_{45}
+\rho_{12}+\rho_{129}+2\rho_{13}+2\rho_{130}-6\rho_{135}-3\rho_{137}-2\rho_{74}
\\\nn
&&
  \rho_{61}=\alpha/2+6\zeta_{20}+3\zeta_{24}-\zeta_{28}-\zeta_{39}-\zeta_{41}-3\zeta_{42}+\zeta_{58}+\rho_{100}+3\rho_{101}-3\rho_{3}
 \\
 &&\rho_{56}= \alpha/2 -\rho_{151}-2\rho_{152}-2\rho_{55},\ \rho_{82}=\alpha/2-\rho_{136}-\rho_{140}-\rho_{76}
\,,
  \rho_{154} =\alpha/4,\ \cdots
\eeqa
where $\cdots$ refer to  some constraints that do not contain $\alpha$. 

Comparing the couplings \reef{cp-1Bp} with the  contact terms in \reef{np-1bp}, one finds the following constraints:  
\beqa
\label{cp-1sm}
&&\tau_6=-\alpha/{4}-\la_{23}+\la_{42}-3\la_{47}+\la_5+\tau_{21},\ \tau_7=\alpha/{8}+\la_{12}/{2}+\la_{19}/{2}+\la_{32}/{2}-\la_{35}/{2}+\la_{10}\nn\\\nn
&& \tau_{17}=-{\alpha}/{4}+\la_{14}+\la_{39},\ \tau_{34}=2\la_{11}-2\la_{17}-\la_{27}+2\tau_{14}-2\tau_{19},\ \tau_{43}=\alpha/{4}+\la_{23}+3\la_{47}\nn\\
&&-\tau_{21}\,,
\tau_{36}={\alpha}/{4}+\la_{23}+3\la_{47}-\tau_{21},\  \tau_{42}=-{\alpha}/{4}+\la_{25}/{2}+\tau_{20}/{2}+\la_{41},\ \cdots
\eeqa
where $\cdots$ refer to the constraints that have no $\alpha$.

Comparing the couplings \reef{cp-1df} with the  contact terms in \reef{np+1bf}, one finds the following constraints:
\beqa
&&\vartheta_{21}=-\alpha/6-\vartheta_{20}/3-\vartheta_{22}/3-\vartheta_{26}/3,\  \vartheta_{30}=\alpha/4\nn\\
&& \pi_{21}=-2\vartheta_{23}+\vartheta_{25}/2-\vartheta_{26}+\pi_{20}/2,\ \cdots
\eeqa
where $\cdots$ refer to the constraints that have no $\alpha$.

Comparing the couplings \reef{cp+1Bf} with the  contact terms in \reef{np-1df}, one finds the following constraints:
\beqa
&\!\!\!\!\!\!\!\!\!\!& \ka_{ 291 }=\alpha/2-\ka_{ 290},\  \ka_{ 333 }=\alpha/4+\gamma_{222}+2\gamma_{224}+3\gamma_{34}+6\gamma_{50}+\iota_{ 219 }+ 2 \iota_{ 222 }- 3 \iota_{ 27 }\\\nonumber
&\!\!\!\!\!\!\!\!\!\!&\ka_{ 334 }=-\alpha/8-\gamma_{222}/2-\gamma_{224}-3\gamma_{34}/2-3\gamma_{50}-\iota_{ 219 }/2-\iota_{ 222 }+3\iota_{ 27 }/2+ 3 \iota_{ 43 }-3\ka_{ 18 }/2\nn\\\nn
&\!\!\!\!\!\!\!\!\!\!&-\ka_{ 321 }-\ka_{ 332}/2,\ \ka_{ 337 }=-\alpha/8-\gamma_{222}/2-\gamma_{224}-\gamma_{225}/2-\gamma_{226}
-3\gamma_{34}/2-3\gamma_{50}-3\gamma_{73}\\\nn
&\!\!\!\!\!\!\!\!\!\!&-3\gamma_{74}/2+3\iota_{ 27 }/2+ 3 \iota_{ 43 }+3\iota_{ 47 }/2+ 3 \iota_{ 55 }-3\ka_{ 10 }/2-3\ka_{ 14 }/2- 3 \ka_{ 19 }-\ka_{ 335 }/2-\ka_{ 336}/2\\\nn
&\!\!\!\!\!\!\!\!\!\!& \ka_{ 338 }=-\alpha/2-2\gamma_{222}-4\gamma_{224}-6\gamma_{34}-12\gamma_{50}- 2 \iota_{ 219 }- 4 \iota_{ 222 }+ 6 \iota_{ 27 }+ 12 \iota_{ 43 }- 6 \ka_{ 10 }+ 2 \ka_{ 317 }\\\nn
&\!\!\!\!\!\!\!\!\!\!&- 2 \ka_{ 335}\,, \ka_{339}=\alpha/4+\gamma_{222}+2\gamma_{224}+3\gamma_{34}+6\gamma_{50}+\iota_{219}+2\iota_{222}-3\iota_{27}
-6\iota_{43}+3\ka_{10}-\ka_{317}\\\nn
&\!\!\!\!\!\!\!\!\!\!&+\ka_{335}\,, \ka_{343}=\alpha/24-\iota_{27}/2,\ \ka_{49}=-\alpha/8-\ga_{27}-\ga_{41}-\iota_{200}/2-\iota_{208}/2+\iota_{25}+\iota_{41}
+\ka_{256}/2\\\nonumber
&\!\!\!\!\!\!\!\!\!\!&\ka_{38}=-\alpha/8-\gamma_{207}+\gamma_{210}/2+\gamma_{24}-\gamma_{41}+\iota_{199}/2-\iota_{208}/2-\iota_{24}+\iota_{41}+3\ka_{270}/2+\ka_{289}+3\ka_{290}\nn\\
&\!\!\!\!\!\!\!\!\!\!& \ka_{39}=\alpha/8-\gamma_{212}/2-\gamma_{216}+\gamma_{27}-3\gamma_{33}/2+\gamma_{41}
-3\gamma_{49}+\iota_{200}/2-\iota_{201}/2+\iota_{208}/2-\iota_{209}-\iota_{25}\nn\\\nonumber
&\!\!\!\!\!\!\!\!\!\!&+3\iota_{26}/2
-\iota_{41}+3\iota_{42}-3\ka_{17}/2-\ka_{297}/2+\ka_{298}/2,\ \ka_{90}=\alpha/24-\ka_{224}\\\nonumber
&\!\!\!\!\!\!\!\!\!\!&\ka_{52}=\alpha/8+\gamma_{212}/2+\gamma_{216}+3\gamma_{33}/2+3\gamma_{49}+\iota_{201}/2+\iota_{209}-3\iota_{26}/2-3\iota_{42}
+3\ka_{17}/2+\ka_{271}\nn\\\nn
&\!\!\!\!\!\!\!\!\!\!&-\ka_{272}/2+\ka_{297}/2-\ka_{298}/2,\ \ka_{70}=\alpha/12-\ga_{19}-2\ga_{36}+\iota_{20}+2\iota_{38}-2\ka_{224}\\\nonumber
&\!\!\!\!\!\!\!\!\!\!&\ka_{6}=\alpha/8+\gamma_{207}-\gamma_{210}/2-\gamma_{24}+\gamma_{41}-\iota_{199}/2+\iota_{208}/2+\iota_{24}-\iota_{41}+\ka_{258}/2
+\ka_{292}/2+\ka_{294}/2
\\\nonumber
&\!\!\!\!\!\!\!\!\!\!& \ka_{342}=-\alpha/8-\gamma_{222}/2-\gamma_{224}-\gamma_{225}/2-\gamma_{226}-3\gamma_{34}/2
-3\gamma_{50}-3\gamma_{73}/2-3\gamma_{74}+3\iota_{27}/2\nn\\
&\!\!\!\!\!\!\!\!\!\!&+3\iota_{43}
+3\iota_{47}/2+3\iota_{55}
-3\ka_{10}/2-3\ka_{14}/2+3\ka_{23}+\ka_{322}-\ka_{335}/2-\ka_{336}/2-3\ka_{340}\nn\\\nonumber
&\!\!\!\!\!\!\!\!\!\!& \ka_{341}=-\alpha/4-\gamma_{222}/2-\gamma_{224}-\gamma_{225}/2-\gamma_{226}-3\gamma_{34}/2
-3\gamma_{50}-3\gamma_{73}/2-3\gamma_{74}+3\iota_{27}/2\nn\\\nn
&\!\!\!\!\!\!\!\!\!\!&+3\iota_{43}+3\iota_{47}/2
+3\iota_{55}-3\ka_{10}/2-3\ka_{14}/2+3\ka_{23}-3\ka_{24}
-\ka_{335}/2-\ka_{336}/2-3\ka_{340}\\\nn
&\!\!\!\!\!\!\!\!\!\!& \ka_{83}=\alpha/12-\ga_{19}-2\ga_{36}+\iota_{20}+2\iota_{38},\  \ka_{90}=\alpha/24-\ka_{224},\ \cdots
\eeqa
where $\cdots$ refer to the constraints that have no $\alpha$.

Comparing the couplings \reef{cp+1hp} with the  contact terms in \reef{cp+1hp0}, one finds the following constraints:
\beqa
&& \psi_{25}=\psi_{26}=
\psi_{31}=\psi_{34}= \psi_{37}= \psi_{50}= \psi_{51}=\psi_{68}=\psi_{69}=\psi_{80}= \psi_{82}=0\\\nonumber
&& \psi_{58}=\alpha-2\psi_{47},\ \cdots
\eeqa
where $\cdots$ refer to the constraints that have no $\alpha$.

Using the fact that there is no contact terms for the couplings of one RR $(p+1)$-form, one dilaton and one scalar field in section 3, one finds the following   constraints on the $\Pi_i$ and $\Lambda_i$ coefficients in \reef{cp+1dp}:
\beqa
&&\Pi_{2}=\Lambda_2\,,\Pi_{4}=\Lambda_4-\Lambda_{10}\,,\Pi_{5}=\Lambda_5-\Lambda_3-\Lambda_4+\Pi_4\,,\Pi_{6}=\Lambda_3\,,\Pi_4=\Lambda_8+\Lambda_3+\Lambda_4\nn\\
&&\Pi_{7}=\Lambda_7-\Lambda_{11}\,,\Pi_{13}=\Lambda_{16}-\Lambda_{22}\,,\Pi_{15}
=\Lambda_{18}-2\Lambda_{22}\,,\Pi_{17}=0\,,\Lambda_{22}=-\Lambda_{24}\labell{PL}
\eeqa

Comparing the couplings \reef{cp+3bp} with the  contact terms in \reef{np+3bf}, one finds the following constraints:
\beqa
&&\t_{18}= \om_{11}= \om_{16}=\om_{18}=\om_{19}= \om_{22}=\om_{28}= \om_{31}= \om_{7}=0\\\nonumber
&&
 \t_{35}=\alpha/{8},\ \t_{87}=-\alpha/{4}-3\t_{12},\ \cdots
\eeqa
where $\cdots$ refer to the constraints that have no $\alpha$.

Imposing the Bianchi identities and ignoring total derivative terms, one finds that the above constraints can not fix all independent coefficients. So one should use another constraint to fix the remaining coefficients. In the next subsection we will use the T-duality constraint to fix the remaining coefficients.

\subsection{T-duality constraint}

The T-duality     transformations on massless field at the leading order of $\alpha'$ are given by the Buscher rules \cite{TB,Meessen:1998qm,Bergshoeff:1995as,Bergshoeff:1996ui,Hassan:1999bv}. The $\alpha'$-correction to these rules have been found in \cite{Bergshoeff:1995cg,Kaloper:1997ux,Bedoya:2014pma} for the Bosonic, Type I and the Heterotic string theories. The Buscher rules in the type II super string theories receive   higher derivative correction (if any) at order $\alpha'^3$ because the first higher derivative correction to the type II supergravities is at eight-derivative level.  In this paper, we are interested in four-derivative couplings on the world-volume of D-branes in type II theories\footnote{One may ask if the T-duality transformations of massless closed string fields   depend on the  presence of D-branes/O-planes. It seems the  answer is no. To see this, we note that, in the type II super string theories, the consistency of  NS-NS couplings at order $\alpha'^2$   on the world volume of  O-plane   with the standard Buscher rules, produces unique couplings which are consistent with S-matrix elements \cite{Robbins:2014ara,Garousi:2014oya}. Similarly, in the bosonic string theory, the consistency of D-brane couplings  at order $\alpha'$ with the Buscher rules and their $\alpha'$-corrections which have been found   in the absence of D-brane \cite{Kaloper:1997ux}, produce correct couplings which are consistent with the S-matrix elements \cite{Garousi:2013gea}.}. As a result, the $\alpha'^3$-corrections of the Bucher rules (if any) do no play any role in our calculations. 

The   Bucher rules are in general nonlinear.  Constraining the world-volume effective actions to be invariant under these nonlinear transformations which may fix all   couplings of bosonic fields,  would be a difficult task. In this paper, however, we are interested  in constraining the world volume couplings of one RR, one NSNS    and one NS   strings  at order $\alpha'^2$ to be invariant under T-duality. Using the fact that the world volume couplings of   one closed and one open strings have no higher derivative corrections in the superstring theory, one realizes that the higher derivative couplings of one RR, one NSNS  and one NS strings   must be invariant under linear T-duality transformations. 

A systematic approach for constructing T-duality
invariant actions is   the   Double Field
Theory \cite{Hull:2009mi,Hohm:2010jy} in which the actions are required to be explicitly invariant under $O(D,D)$ transformations. The modification of this theory to Double $\alpha'$-geometry in which the
generalized Lie derivative receives $\alpha'$- corrections,  requires and determines the higher derivative couplings \cite{Hohm:2013jaa,Marques:2015vua}. Our   approach, however, is that the actions are required to be invariant under the Buscher rules and their $\alpha'$-corrections. Since in type II superstring there is no $\alpha'^2$-corrections to the Buscher rules, we require the D-brane effective action at order $\alpha'^2$ to be invariant under the standard Buscher rules. In the particular case of two closed and one open strings in which we are interested, the couplings must be invariant under the linearized Buscher rules as well. 
We refer the interested reader to, for example, \cite{Jalali:2015xca} for the list of liner T-duality transformations\footnote{ Massless world-volume   fields may receive $\alpha'$-correction. Since D$_p$-brane along the Killing direction transforms under T-duality to  D$_{p-1}$-brane, the general form of the T-duality transformation of the world volume  gauge field along the Killing direction $y$ is $A_y\rightarrow f(\phi^y,\prt\phi^y, \prt\prt\phi^y,\cdots)$ where $f$ at $\alpha'^0$ is $\phi^y$. At order $\alpha'$, it may be $a\alpha'\phi^y\prt\phi^y\prt\phi^y+b\alpha'\prt\prt\phi^y$ where $a,b$ are constants. Consistency of S-matrix elements of open strings with T-duality Ward identity dictates that the linear term is zero, \ie $b=0$. The coefficient of the nonlinear term may be non-zero, however, this term  play no role for the couplings of two closed and one open string in which  we are interested in this paper. Similarly for the possible corrections at higher order of $\alpha'$.} and for the method to constrain the couplings to be invariant under the linear T-duality.

To impose the T-duality constraint on the world volume action $S_p$, we have to consider all couplings in section 4.1 and the couplings of two closed and one open strings that are resulted from the Taylor expansion and pull-back operator in \reef{LTdual} as well as the couplings of two closed and one open strings in \reef{CS2f}. Including all these couplings in $S_p$ and imposing the S-matrix constraints found in the previous section on $S_p$, then the action must be invariant under linear T-duality\footnote{In fact we have checked that the Feynman amplitudes in sections 3.1-3.5 satisfy the T-dual Ward identity. As a result, the couplings in S$_p$ must also satisfy the T-dual Ward identity because the S-matrix element of one RR, one NSNS and one NS states at order $\alpha'^2$ which includes the couplings in S$_p$ and the Feynman amplitudes must satisfy the Ward identity.   }. This produces some new constraints.

We begin by  imposing the T-duality constraint on the couplings \reef{cp-3bf}.  Concerning the indices of the RR field strength, there are  two cases to consider. The Killing index is carried either by the RR field strength or by the NSNS and NS fields. We have found that the couplings \reef{cp-3bf} are invariant under T-duality when the Killing index is carried by the RR field.  However, 
 when the Killing index is carried by the NSNS and NS fields, the T-duality transforms the RR $(p-3)$-form to the RR $(p-1)$-form. 

The couplings involving the RR $(p-1)$-form, must be invariant under T-duality when the Killing index is carries by the RR field. This produces the following constraints:  
\beqa
&&\!\!\!\!\!\!\!\!\la_{46}=\alpha/4+\la_{42}/3,\ \rho_{138}=\alpha/2-\rho_{129}-2\rho_{130}-\rho_{131}/2-\rho_{132}+6\rho_{135}+\rho_{136}+3\rho_{137}\label{cp-1st}\\\nn
&& \!\!\!\!\!\!\!\!\rho_{132}=-\alpha/2+4\delta_{11}+4\delta_{16}+4\delta_{19}-\zeta_{15}-2\zeta_{23}-\zeta_{26}
-\zeta_{35}+\zeta_{36}-2\la_{42}+\rho_{12}+2\rho_{13}-\rho_{131}/2\\\nn
&&\!\!\!\!\!\!\!\!\la_{47}=-\alpha/12+\la_{42}/3,\  \rho_{144}=\alpha/4-2\delta_{11}-2\delta_{16}-2\delta_{19}+\zeta_{15}/2+\zeta_{23}+\zeta_{26}/2+\zeta_{35}/2-\zeta_{36}/2\nn\\
&&\!\!\!\!\!\!\!\!-\rho_{12}/2-\rho_{13}-\rho_{133}/2-3\rho_{135}-\rho_{136}/2-3\rho_{137}/2+3\rho_{139}/2-\rho_{140}/2
\nn\\&&
\!\!\!\!\!\!\!\! \pi_8=\alpha/2-\z_{17}-3\z_{42}-3\vartheta_1-\vartheta_{11}+\vartheta_{12}+3\vartheta_2+3\pi_1+\pi_3-3\pi_5,\ \cdots
\eeqa 
where $\cdots$ refer to the constraints that have no $\alpha$. Imposing the above constraint in $S_p$, one finds that when the Killing index is carries by the NSNS and NS fields in \reef{cp-3bf}, they are transform to the couplings involving the  RR $(p-1)$-form after imposing  the following constraints:   
\beqa
&&\!\!\!\!\!\!\!\! \nu_{2}=-\alpha/6-2\delta_{10}/3+\delta_{12}/3+\delta_3+\delta_{34}/3+\delta_{37}/3+\delta_7+2\mu_{12}/3+\mu_{2}+\mu_{33}/3+\nu_{15}/3-2\nu_{16}/3
 \nn\\\nonumber
&&\!\!\!\!\!\!\!\!
\nu_{31}=-\alpha/8-\delta_{13}/2+\delta_{14}-\delta_{15}/2-\delta_{16}+\delta_{18}-\delta_{19}+\mu_{12}-\mu_{15}/2
+\mu_{16}-\mu_{17}/2+\mu_{20}+\nu_{15}/2\nn\\
&&\!\!\!\!\!\!\!\!-\nu_{16}+\nu_{30}/2,\ \tau_{15}=-\alpha/8+\delta_{11}+\delta_{16}+\delta_{19}-\la_{10}-\la_{12}/2-\la_{19}/2+\la_{32}/2+\la_{35}/2
\nn\\\nn
&&\!\!\!\!\!\!\!\!\tau_{29}=-\alpha/4+2\delta_{10}-\delta_{12}-\delta_{13}+2\delta_{14}-\delta_{15}+2\delta_{18}-3\delta_2-\delta_{29}-3\delta_3-\delta_{34}-\delta_{37}-3\delta_7\nn\\
&&\!\!\!\!\!\!\!\!
-2\la_{10}-\la_{12}-3\la_{18}
-\la_{19}-2\la_{2}+\la_{31}+\la_{32}+\la_{35}-3\tau_{1}
\nn\\\nonumber
&&\!\!\!\!\!\!\!\!\nu_{45}=-\alpha/8-\delta_{41}/2+\delta_{43}-\delta_8/2+\delta_9+\mu_{28}/2+3\mu_{52}/2+\mu_{59}/2+3\mu_{60}/2+\nu_{22}/2\nn\\
&&\!\!\!\!\!\!\!\! \la_{4}=-\alpha/4-2\la_{10}-\la_{12}-\la_{19}-2\la_{2}+\la_{31}+\la_{32}\\\nn
&&\!\!\!\!\!\!\!\! \tau_{46}=-\alpha/12-\delta_{25}/3-\delta_{41}/3+2\delta_{43}/3-\delta_{45}-\delta_8/3+2\delta_9/3+\la_{24}/3+2\la_{40}/3-\tau_{22}/3
\\\nonumber
&&\!\!\!\!\!\!\!\! \nu_{4}=-\alpha/8+3\mu_{1}+\mu_{12}+\mu_{34}-\mu_{35}/2-3\mu_{5}/2+3\nu_{1}
+\nu_{10}+\nu_{15}/2-\nu_{16}-3\nu_{20}/2+\nu_{23}/2\nn\\
&&\!\!\!\!\!\!\!\!-\nu_{27}/2-\delta_{10}-\delta_{11}+\delta_{12}/2-\delta_{16}-\delta_{19}+\delta_{35}/2-\delta_{36}
+\delta_{37}/2-3\delta_4+3\delta_6/2+3\delta_7/2+\zeta_{15}/4\nn\\\nn
&&\!\!\!\!\!\!\!\!-\zeta_2/4+\zeta_{23}/2
+\zeta_{26}/4+\zeta_{35}/4-\zeta_{36}/4-\zeta_5/2,\ \cdots
\eeqa
where $\cdots$ refer to the constraints that have no $\alpha$.

Imposing the above constraints in $S_p$, one finds the couplings involving RR $(p-3)$-form are invariant and the couplings involving RR $(p-1)$-form are invariant when the Killing index is carried by the RR field. Otherwise, they transform to the couplings involving RR $(p+1)$-form after imposing the following constraints: 
\beqa
&&\ka_{332}=-\alpha/4+3\iota_{27}+6\iota_{43}-3\ka_{18}
,\ \la_{2}=-\alpha/8-\la_{10},\ \ka_{335}=-\alpha/4+3\iota_{27}+6\iota_{43}-3\ka_{10}\nn\\
&&
 \rho_{140}=-\alpha/2-6\rho_{135}-\rho_{136}
-3\rho_{137},\ \rho_{137}=-\alpha/6-2\rho_{135}
\\\nn
&&
 \rho_{13}= \alpha/4+\zeta_{15}/2+\zeta_{23}+\zeta_{26}/2+\zeta_{35}/2-\zeta_{36}/2-\rho_{12}/2
\\\nonumber
&&\iota_{79}=-\alpha/8+\iota_{25}+\iota_{41},\ \psi_3=-\alpha/4-\psi_{13}/3,\ \cdots
\eeqa
where $\cdots$ refer to the constraints that have no $\alpha$. The   couplings involving RR $(p+1)$-form must also be invariant under T-duality when the Killing index is carried by the RR field. This produces the following constraints:
\beqa
\Pi_{16}=\Lambda_{23}=0\,,\,\iota_{200}=\iota_{194}-\iota_{196}/2-\iota_{199},\ \cdots
\eeqa
which have no $\alpha$.

Imposing the above constraints in $S_p$, one finds the couplings involving RR $(p-3)$-form and RR $(p-1)$-form are invariant and the couplings involving RR $(p+1)$-form are invariant when the Killing index is carried by the RR field. Otherwise, they transform to the couplings involving RR $(p+3)$-form. The latter couplings are invariant under T-duality when the Killing index is carried by the RR field provided that  
\beqa
&&\omega_{17}=0,\ \theta_{79}=\theta_{19}
\eeqa
Imposing these constraints on the $(p+3)$ couplings, one finds  when the Killings index in the couplings involving RR $(p+1)$-form is carried by NSNS and NS fields, they transform to the $(p+3)$ couplings after imposing the following constraints: 
\beqa
\label{cp+1t}
&& \omega_{3}= \omega_{58}= \omega_{83}= \omega_{12}=\omega_4=\iota_{206}=0,\ \omega_{14}=-\alpha/8\\\nn
&&\theta_{17}=-\alpha/8,\ \theta_{85}=\alpha/4-3\theta_{4},\ \cdots
\eeqa
where $\cdots$ refer to the constraints that have no $\alpha$. Imposing all  above constraints in $S_p$, one finds the couplings involving RR $(p-3)$-form, RR $(p-1)$-form, RR $(p+1)$-form and RR $(p+3)$-form are invariant under linear T-duality.

After imposing all constraints on the couplings in section 4.1, we have found  two sets of couplings. One set is the couplings which have coefficient $\alpha$. 
They involve  
the following couplings for $\cF^{(p-2)}$:
\beqa
S_{p}^{WS} &\!\!\!\!\!\supset\!\!\!\!\!&-\frac{\pi^2\alpha'^2T_{p}}{24}\frac{1}{4}\int d^{p+1}x\, \epsilon^{a_0a_1a_2a_3 a_4}\bigg[H_{aba_2} \
\prt^{a}{}\tB_{a_0a_1} \prt^{b}\cF^{(2)}_{a_3a_4}\nn\\
&&-2H_{aa_2}{}_{i}\prt^{a}{}\tB_{a_0a_1}\prt^{i}{\cF}^{(2)}_{a_3a_4}-H_{ia_1a_2}\prt^{a}\tB_{aa_0}\prt^{i}{\cF}^{(2)}_{a_3a_4 }\nn\\
&&- H_{ba_1a_2}\prt_{a_0}\tB^{ab}\prt_{a}{\cF}^{(2)}_{a_3a_4 }+2\tB_{a_0a_1}\prt_{a_2}H^{ic}{}_{a_3}{}\prt_{i}{\cF}^{(2)}_{ca_4}\nn\\
&&+2H^i{}_{a_2a_3}\prt_{a_1}{}\tB_{aa_0}\prt^{a}{\cF}^{(2)}_{ia_4 }+4H^{b}{}_{a_2a_3}\prt_{a_1}\tB_{aa_0}\prt^{a}{\cF}^{(2)}_{ba_4 }\nn\\
&&+H_{ca_1a_2}\prt^{c}\tB_{aa_0}\prt^{a}{\cF}^{(2)}_{a_3a_4 }-H_{ba_1a_2}\prt^a\tB_{aa_0}\prt^b\cF^{(2)}_{a_3a_4}\nn\\
&&
+H_{aa_2a_3}\prt^b\tB_{a_0a_1}\prt^a\cF^{(2)}_{ba_4 }+\frac{2}{3}\bigg(3\tB_{a_0a_1}\prt_{i}H^{c}{}_{a_2a_3}\prt^{i}{\cF}^{(2)}_{ca_4 }\nn\\
&&+3\tB_{a_0a_1}\prt^{c}H^i{}_{a_3a_4}\prt_{a_2}{\cF}^{(2)}_{ic }-2\tB^a{}_{a_0}\prt_{b}H_{a_1a_2a_3}\prt^{b}{\cF}^{(2)}_{aa_4 }\nn\\
&&+4H_{a_1a_2a_3}\prt^{a}\tB_{aa_0}\prt^{c}{\cF}^{(2)}_{ca_4 }-4H_{a_1a_2a_3}\prt^{c}\tB^{a}_{}{a_0}\prt_{c}{\cF}^{(2)}_{aa_4 }\bigg)\bigg]
\eeqa
where $p=4$, 
 the following couplings for $\cF^{(p)}$: 
\beqa
S_{p}^{WS} &\!\!\!\!\!\supset\!\!\!\!\!&-\frac{\pi^2\alpha'^2T_{p}}{24}\frac{1}{3!}\int d^{p+1}x\, \epsilon^{a_0a_1a_2 a_3}\bigg[\prt_{j}{\cF}^{(3)}_{a_1a_2 a_3}H_{ai}{}^{j}\Omega_{a_0}{}^{ai}-\frac{3}{2!}\,\prt_{b}{\cF}^{(3)}_{ia_2 a_3}H^{b}{}_{a_0a_1}\Omega_{a}{}^{ai}\nn\\
&&-3\cF^{(3)}_{ia_2 a_3}\Omega_{a_0}{}^{ai}\prt^{b}H_{aba_1}
-\frac{3}{2!}\cF^{(3)}_{ia_2 a_3}\Omega_{a}{}^{ai}\prt^{b}H_{ba_0a_1}+\frac{3}{2!}\cF^{(3)}_{ia_2 a_3}\Omega^{bai}\prt_{b}H_{aa_0a_1}\nn\\
&&+\cF^{(3)}_{ija}\prt^jH_{a_0a_1a_2}\Omega_{a_3}{}^{ai}+3 \tB_{a_0a_1}\cR^{ij}\prt_{j}{\cF}^{(3)}_{ia_2 a_3}-2\tB_{aa_0}\cR^{ai}\prt_{i}{\cF}^{(3)}_{a_1a_2 a_3}\nn\\
&&-6 \tB^{ab}\cR_{ba_1}\prt_{a_0}{\cF}^{(3)}_{aa_2 a_3}+6 \tB_{aa_0} R^{ai}{}_{ba_1}\prt^{b}{\cF}^{(3)}_{ia_2 a_3}-2 \tB^{ab}R_{aa_0bi} \prt^{i}{\cF}^{(3)}_{a_1a_2 a_3}\nn\\
&&+6 \tB_{a_0a_1} R_{aija_2}\prt^{a}{\cF}^{(3)}_{ija_3 }-6\tB^{a}{}_{a_0}R^{i}{}_{a_1a_2}{}^{j}\prt_{a}{\cF}^{(3)}_{ija_3 }\nn\\
&&+3 \tB_{ab} R^{bi}{}_{a_0a_1}\prt^{a}{\cF}^{(3)}_{ia_2 a_3}
-12\tB^{a}{}_{a_0}R^{aij}{}_{a_1}\prt_{a_2}{\cF}^{(3)}_{ija_3 }\bigg]
\eeqa
where   $p=3$, 
the following couplings for $\cF^{(p+2)}$: 
\beqa
 S_{p}^{WS} &\!\!\!\!\!\supset\!\!\!\!\!&-\frac{\pi^2\alpha'^2T_{p}}{48}\frac{1}{3!}\int d^{p+1}x\, \epsilon^{a_0a_1 a_2}\bigg[-24{\cF}^{(4)}_{ijka_2 }R^{bjk}{}_{a_1}\Omega_{ba_0}{}^{i}\nn\\
&&-6{\cF}^{(4)}_{ijka_2 } R_{a_0a_1}{}^{jk}\Omega_{a}{}^{ai} 
+3\tB_{a_0a_1}\prt^{k}H^{aij}\prt_{a}{\cF}^{(4)}_{ijka_2}\nn\\
&&+3\tB_{a_0}{}^{b}\prt_{b}H^{aij}\prt_{a}{\cF}^{(4)}_{ija_1 a_2}
-3\tB_{a_0}{}^{b}\prt_{a}H^{aij}\prt_{b}{\cF}^{(4)}_{ija_1 a_2}\nn\\
&&
+3\tB_{a_0}{}^{b}\prt^{a}H_{b}{}^{ij}\prt_{a}{\cF}^{(4)}_{ija_1 a_2}
+6\tB_{a_0}{}^{b}\prt^{k}H_{a_1}{}^{ij}\prt_{b}{\cF}^{(4)}_{ijka_2}\nn\\
&&
-6\tB_{a_0}{}^{b}\prt^{k}H_{b}{}^{ij}\prt_{a_1}{\cF}^{(4)}_{ijka_2}-3 H^{bij} \prt_{a}\tB^{a}{}_{a_0}\prt_{b}{\cF}^{(4)}_{ija_1a_2}\nn\\
&&+2\tB^{ab}\prt_{b}H_{a}{}^{ci}\prt_{c}{\cF}^{(4)}_{ia_0a_1 a_2}
-2\tB^{ab}\prt_{f}H_{a}{}^{fi}\prt_{b}{\cF}^{(4)}_{ia_0a_1 a_2}
-\tB^{ab}\prt^{f}H_{ab}{}^{i}\prt_{f}{\cF}^{(4)}_{ia_0a_1 a_2}\nn\\
&&
-\tB^{ab}\prt^{j}H_{ab}{}^{i}\prt_{i}{\cF}^{(4)}_{ja_0a_1 a_2}
-\tB^{ab}\prt^{j}H_{ab}{}^{i}\prt_{j}{\cF}^{(4)}_{ia_0a_1 a_2}
+3\tB^{ab}\prt_{a_0}H_{a}{}^{ij}\prt_{b}{\cF}^{(4)}_{ija_1  a_2}\bigg]
\eeqa
 where  $p=2$, 
and the following couplings for $\cF^{(p+4)}$: 
 \beqa
S_{p}^{WS} &\!\!\!\!\!\supset\!\!\!\!\!&\frac{\pi^2\alpha'^2T_{p}}{48}\frac{1}{2}\int d^{p+1}x\, \epsilon^{a_0a_1 }\bigg[{\cF}^{(5)}_{ijka_0a_1}\bigg(
\Omega_{a}{}^{ai}\prt_{b}H^{bjk}-\Omega^{abi}\prt_{a}H_{b}{}^{jk}
\bigg)\nn\\
&&
+2\cF^{(5)}_{ijkla_1}\bigg(\Omega^{c}{}_{a_0}{}^{i}\prt^{l}H_{c}{}^{jk}
- \Omega_{c}{}^{ci}\prt^{l}H_{a_0}{}^{jk}\bigg)+H^{bjk}\Omega_{a}{}^{ai}\prt_{b}{\cF}^{(5)}_{ijka_0a_1}\bigg]
\eeqa
where  $p=1$. 

The above results can easily be extended to   arbitrary $p$ by requiring that each term must be invariant under linear T-duality when the Killing index is carried by the RR field strength, \eg D$_4$-brane coupling $\frac{1}{2}\epsilon^{a_0a_1a_2a_3 a_4}H_{aba_2} 
\prt^{a}{}\tB_{a_0a_1} \prt^{b}\cF^{(2)}_{a_3a_4}$ is extended to the following D$_p$-brane coupling:
\beqa
  \frac{1}{(p-2)!}\epsilon^{a_0\cdots a_p}H_{aba_2} 
\prt^{a}{}\tB_{a_0a_1} \prt^{b}\cF^{(p-2)}_{a_3\cdots a_p}\labell{Dpb}
\eeqa
  When the world volume Killing index $y$ is carried by the RR, it becomes
\beqa
\frac{1}{(p-3)!}\epsilon^{a_0\cdots a_{p-1}y}H_{aba_2} 
\prt^{a}{}\tB_{a_0a_1} \prt^{b}\cF^{(p-2)}_{a_3\cdots a_{p-1}y}\nn
\eeqa
Under T-duality D$_p$-brane transforms to D$_{p-1}$-brane and the above coupling transforms to 
\beqa
\frac{1}{(p-3)!}\epsilon^{a_0\cdots a_{p-1}}H_{aba_2} 
\prt^{a}{}\tB_{a_0a_1} \prt^{b}\cF^{(p-3)}_{a_3\cdots a_{p-1}}\nn
\eeqa
which is the same as the coupling \reef{Dpb} for D$_{p-1}$-brane. Performing similar extensions for all other couplings, one finds the couplings   in \reef{cp-3bfaction}, \reef{p-1hfaction}, \reef{act1}  and \reef{p+3bpaction}.  

Another set is the couplings which have unfixed coefficients. However, they all are canceled after writing the field strengths in terms of field potentials and ignoring total derivative terms. That means, up to total derivative terms and the Bianchi identities, the couplings in \reef{cp-3bfaction}, \reef{p-1hfaction}, \reef{act1}  and \reef{p+3bpaction} are the unique couplings which are consistent with the contact terms of the S-matrix element at order $\alpha'^2$ and are consistent with  the linear T-duality.



{\bf Acknowledgments}:   This work is supported by Ferdowsi University of Mashhad.

\end{document}